\documentclass[11pt,twoside]{article}

\usepackage{epsfig,amsfonts,color}
\usepackage{cite}
\usepackage{mcite}
\usepackage{amsmath}
\usepackage{mathtools}

\usepackage[outdir=./]{epstopdf}
\usepackage{amssymb, palatino, geometry,url}
\usepackage{amsthm}
\usepackage{algorithmic}
\usepackage{booktabs}
\usepackage{siunitx}
\usepackage[noresetcount,lined,boxed]{algorithm2e} 

\usepackage[colorlinks=true,linkcolor=blue,citecolor=blue,urlcolor=blue]{hyperref}
\usepackage{subcaption}
\usepackage{cleveref}
\usepackage{float}
\usepackage{fancyhdr}
\theoremstyle{definition}

\newcommand{\be}{\begin{equation}}
\newcommand{\ee}{\end{equation}}
\newcommand{\bea}{\begin{eqnarray}}
\newcommand{\eea}{\end{eqnarray}}

\newcommand{\bvec}{\left(\begin{array}{c}}
\newcommand{\evec}{\end{array}\right)}
\newcommand{\bsub}{\begin{subequations}}
\newcommand{\esub}{\end{subequations}}

\usepackage[thinc]{esdiff}

\usepackage{lineno}

\title{Topological Data Analysis for Particulate Gels}

\author{Alexander Smith${}^{\text{a}}$, Gavin J. Donley${}^{\text{b}}$, Emanuela Del Gado${}^{\text{b,*}}$, and Victor M. Zavala${}^{\text{c,d,*}}$\\
{\small ${}^{\text{a}}$Department of Chemical Engineering and Material Science}\\
{\small \;University of Minnesota, Minneapolis, MN, USA}\\
{\small ${}^{\text{b}}$Department of Physics}\\
{\small \;Georgetown University, Washington, DC, USA}\\
{\small ${}^{\text{c}}$Department of Chemical and Biological Engineering}\\
{\small \;University of Wisconsin-Madison, Madison, WI, USA}\\
{\small ${}^{\text{d}}$Mathematics and Computer Science Division}\\
{\small \;Argonne National Laboratory, Lemont, IL, USA}\\
{\small ${}^{*}$Corresponding Authors:  \href{mailto:ed610@georgetown.edu}{ed610@georgetown.edu},  \href{mailto:victor.zavala@wisc.edu}{victor.zavala@wisc.edu}}
}

\begin{document}

\maketitle

\begin{abstract}
Soft gels, formed via the self-assembly of particulate organic materials, exhibit intricate multi-scale structures that provides them with flexibility and resilience when subjected to external stresses. This work combines molecular simulations and topological data analysis (TDA) to characterize the complex multi-scale structure of soft gels. Our TDA analysis focuses on the use of the Euler characteristic, which is an interpretable and computationally-scalable topological descriptor that is combined with filtration operations to obtain information on the geometric (local) and topological (global) structure of soft gels. We reduce the topological information obtained with TDA using principal component analysis (PCA) and show that this provides an informative low-dimensional representation of gel structure. We use the proposed  computational framework to investigate the influence of gel preparation (e.g., quench rate, volume fraction) on soft gel structure and to explore dynamic deformations that emerge under oscillatory shear in various response regimes (linear, nonlinear, and flow). Our analysis identifies specific scales and extents at which hierarchical structures in soft gels are affected; moreover, correlations between structural deformations and mechanical phenomena (such as shear stiffening) are explored. In summary, we show that TDA facilitates the mathematical representation, quantification, and analysis of soft gel structures, extending traditional network analysis methods to capture both local and global organization.
\end{abstract}

\section{Introduction}

Particulate gel technology has emerged as a versatile and innovative solution with widespread applications across diverse industries, including pharmaceuticals, foods, and construction. In the pharmaceutical and medical device industries, these gels play a pivotal role, serving as effective drug delivery vehicles, encapsulation systems, and scaffolds. Their deformability enables the seamless incorporation of various compounds, medications, or diagnostic agents, contributing to enhanced bioavailability and therapeutic efficacy in patient care \cite{burey2008hydrocolloid, tadros2015interfacial, livage1982transition}. Within the food industry, soft gel encapsulation is employed for the delivery of functional ingredients, vitamins, and supplements. The encapsulation of sensitive compounds not only enhances their stability but also facilitates precise dosage control, meeting the growing consumer demand for convenient and palatable nutritional solutions \cite{cao2020design, dickinson2015microgels, townsend2019flow, wang2012comparison}. In the construction sector, the utility of soft gels extends to the realm of binders, sealants, adhesives, and coatings, contributing to improved performance and durability of building materials \cite{zhang2011building, yan2016review, Ioannidou2016c, liu2017mitigation}.
\\

Gel formation may proceed through phase separation, aggregation, and self-assembly of polymers, colloids, or other soft matter components \cite{bouzid2018network}. The stochastic and multi-scale nature of these mechanisms produces hierarchical, amorphous structures, and provides  texture, elasticity, and stability \cite{bouzid2018network, bouzid2020mechanics, lieleg2011slow, conrad2010arrested, koumakis2015tuning}. The multi-scale structure of gels can be represented mathematically as networks defined by the interactions of their building blocks (e.g., particles, droplets, polymer aggregates). These network representations provide insight into the deformation of soft gels, and aid in understanding their mechanical failure \cite{bouzid2018network, bouzid2020mechanics, lieleg2011slow, burla2020Connectivity,nabizadeh2024network}. For example, while soft gels can accommodate large deformations, micro-scale cracks that result from the accumulation of local tension can nucleate and trigger macroscopic failure, in susceptible points determined by both the local spatial distribution (geometry) and the large scale connectivity (topology) of the gel network \cite{bouzid2018network, koumakis2015tuning}. The network structures are not static but highly dynamically re-configurable; deformations, such as squeezing or stretching, and deformation rates can modify the mechanical characteristics of a gel by altering its network structure \cite{koumakis2015tuning, bouzid2018network, bouzid2020mechanics, colombo2014stress, manley2005time, Sudreau2023shear}. In addition, relaxation of the stresses accumulated in the gel through the gel formation itself, under different environmental conditions, is another important source of restructuring \cite{bouzid2018network, bouzid2020mechanics, colombo2014stress}. 
\\

The adaptable nature of particulate gel materials, their flexibility and tunability, provide exciting opportunities to precisely control and design both geometry {\it and} topology of the particle networks, potentially opening up a broad space for metamaterials design and discovery \cite{moghimi2017colloidal,gibaud2020rheoacoustic,bonn2017yield}. 
The geometry of the structure (its local properties) can be controlled through the physical chemistry of particles and solvent, particles surface properties, or depletion interactions \cite{Israelachvili2011}, and, in certain cases, is directly accessible through confocal microscopy imaging\cite{dinsmore2006microscopic,Whitaker2019,royall2021real}. However, getting access to the large-scale (global) organization of the full network and its topology remain a challenge. Computer simulations can provide this type of access, but many of the methods used in understanding the structure of gels focus exclusively on geometry or topology, but not both. From a geometric perspective, the computation of the particle radial distribution functions provides a link to scattering experiments and constitutes an excellent tool to quantify the local structure \cite{hsiao2012role, ohtsuka2008local, varadan2003direct} but does not capture the topology of the network \cite{smith2021euler, mantz2008utilizing, jiao2007modeling}. Voronoi and Delaunay tesselation can capture properties such as cavity size distribution \cite{arizzi1992space, hansen2001structural, lotito2020pattern}, but provide limited quantification tools of the topology of a gel structure.
\\

Graph-theoretic methods use graph descriptors (e.g., modularity, average path length, bond number, and minimal cycle basis) to quantify the structure of a network \cite{bouzid2018network,karnes2020network,nabizadeh2024network}. These descriptors, however, are based on averages of local topological structures and thus might fail to capture the multi-scale nature of soft gels. Parametric methods based for example on distance thresholds or $k$-nearest neighbors commonly used to identify a graph \cite{glielmo2021unsupervised} provide results which are often highly susceptible to the selection of parameter (e.g., distance threshold, $k$-neighbors), which is problematic if the parameter selection is not obvious from the data and rooted in the physics of the system. It is also worth noting that some of these graph-theoretic methods are computationally expensive. For example, the identification of a minimal cycle basis for a graph is at least polynomial in the number of nodes and edges present \cite{kaveh1994revised}, requiring significant computational time for simulations of soft gel structures from simulations which contain hundreds of thousands of nodes and edges to be able to reproduce the heterogeneities typical of real materials. It is also important to highlight that a network/graph is inherently a 2-dimensional object; as such, these representations might miss important information of the 3-dimensional structure of soft gels.
\\

In this work, we propose an approach, based on molecular simulations and topological data analysis (TDA), that captures both the geometry {\it and} topology of the 3-dimensional structure of soft particulate gels. Specifically, we focus on the application of the Euler characteristic (EC) \cite{smith2021euler, smith2022topological}, which is an intuitive, interpretable, and computationally-scalable topological descriptor. This approach quantifies the structure of the gel network using basic topological invariants such as the number of connected components, cycles (holes), and voids. We use a mathematical technique known as a filtration to quantify how topological invariants emerge and disappear at different length scales. This information is condensed in a topological summary known as the EC curve. These approaches have been successfully applied in the analysis of complex materials, simulations, and flow networks \cite{rocks2021hidden,rocks2020revealing, smith2021topological, smith2021euler, smith2022topological, chung2021reviews, obayashi2022persistent, nakamura2015persistent}. The approaches allow us to bypass the need for parameter selection, improving the robustness of our analysis in comparison to other parametric methods (e.g., distance thresholding, neighbor selection). Moreover, these approaches allow us to capture both topological (global) and geometrical (local) characteristics of soft gel structures \cite{mecke1991euler, klee1963euler}.   
\\

We find that our proposed computational framework effectively represents, quantifies, and summarizes the multi-scale nature of gel structures obtained from molecular simulations. We reduce the topological information obtained with TDA using principal component analysis (PCA) and show that this provides an informative low-dimensional representation of gel structure. The application of the proposed computational framework reveals the influence of variations of the gelation kinetics on the local organization of the gels, alongside with the impact of volume fraction changes on both local and global structural aspects. Our analysis also identifies topological transitions underpinning the onset of nonlinear response and flow when the gels are subjected to large amplitude oscillatory tests, clearly disentangling recoverable/unrecoverable changes from damage accumulation and revealing their hierarchical nature; this ultimately reveals a connection between the topological nature of the soft gel material properties and their rheological response. The methods presented here are computationally scalable (analysis of large simulations can be done in seconds-minutes and on a personal computer), interpretable, and require minimal data. The proposed approach can also be used for analysis of experimental data (such as confocal imaging), providing a route for a more direct comparison of molecular simulation and experiments \cite{dinsmore2002direct, dong2022direct, habdas2002video}. All data and scripts needed for reproducing the results are shared as open-source code.  

\section{Results}

\begin{figure}[!htp]
     \centering
     \includegraphics[width=0.95\textwidth]{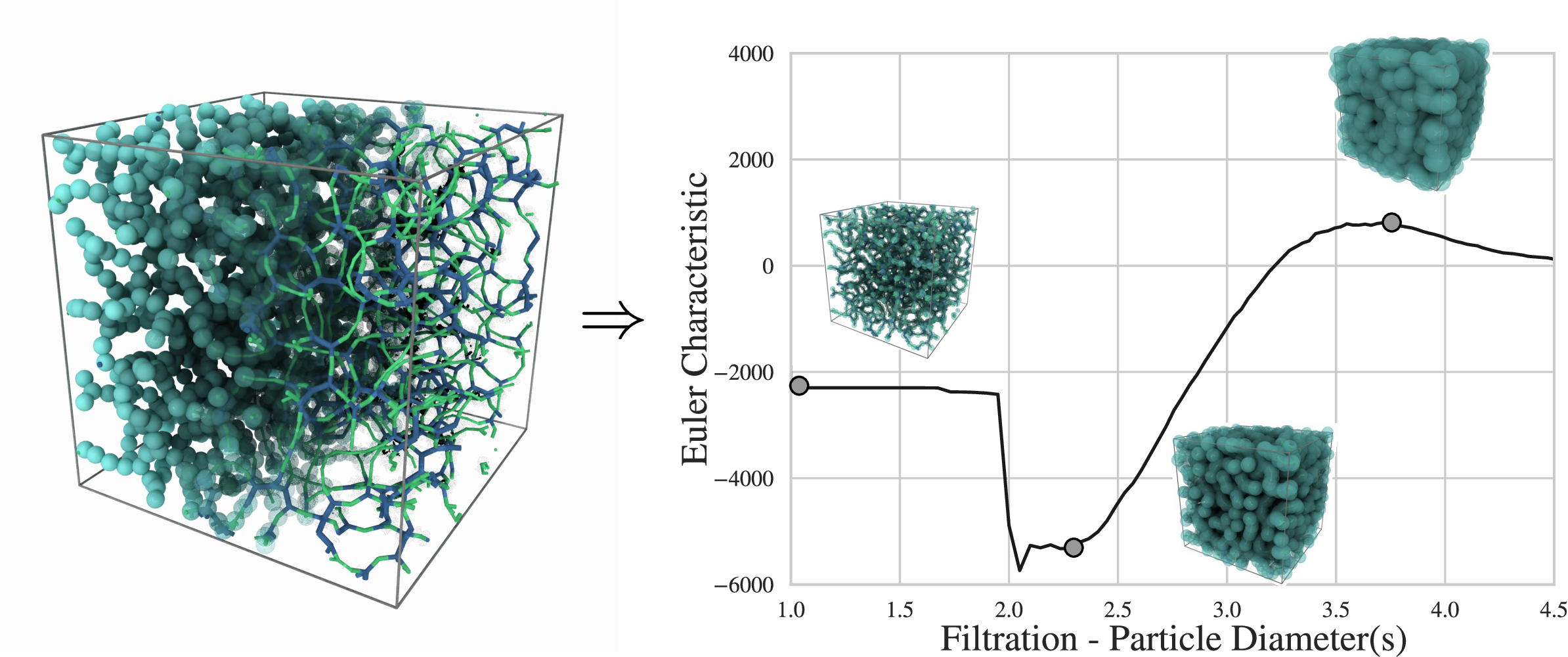}
     \caption{(left) Visualization of a soft gel for a single simulation frame. We provide this visualization to illustrate how the various topological features: components, holes, voids, could evolve during a filtration. (right) EC curve for the gel simulation along with a visualization of the changes in the simulation topology at various filtration diameters. The EC curve captures various topological characteristics found at multiple scales within the soft gel and show how the topological features such as holes ($\sim$2.3 particle diameters) and large voids ($\sim$3.7 particle diameters) emerge. We also note that there is minimal change in topology between $D = 1$ and $D = 2$ particle diameters. This stable topology reflects the bonded structure of the gel, and is often what is studied via network analysis tools. As we pass to $D > 2$ particle diameters, we see large topological transitions in the material, which now reflects the 3-dimensional structure of the gel and captures overlap between particles that are not necessarily directly bonded.}
     \label{fig:3dfilt}
\end{figure}

To understand and quantify the structure of particulate gel simulations, we combine molecular simulations with the analysis of topological invariants of the gel structures based on the EC, filtration operations, and PCA. These methods are combined to represent and quantify the gel structures across multiple length scales and capture the changes they undergo during shear. PCA is used to visualize topological changes and provides insight into the impact of diverse drivers on gel response (such as strain stiffening/hardening or yielding). A summary of the computational workflow to quantify the hierarchical topology of a soft gel is presented in Figure \ref{fig:3dfilt}. 

\subsection{Topological Analysis of Molecular Simulations} \label{topo_anal}

We use a particulate gel model that captures most aspects of microscopic dynamics and rheology of this class of materials. Each gel is composed of $N$ particles, which interact via attractive short-range interactions of maximum strength $\epsilon$, mediated by the solvent in which they are immersed and through which their thermal motion is overdamped. Surface roughness, shape irregularity and sintering processes can limit the relative motion of particles as they aggregate in real materials. These effects are included in the model through an angular modulation of the net attraction that introduces a bending rigidity of the interparticle bonds \cite{colombo2014stress,bouzid2017elastically,bouzid2020mechanics,bantawa21,bantawa23}. We use a cubic box of size $L$ with periodic boundary conditions and number density $N/L^3$ which corresponds to an approximate solid volume fraction $\phi =\frac{N \pi d^3/6}{(Ld)^3}$. For each value of $\phi$, various gel microstructures are obtained by tuning the rate $\Gamma$ at which the relative strength of the attractive interactions (with respect to $k_{B}T$) is increased to induce gelation during the sample preparation (see Methods). Finally, we subject a subset of the gels to large amplitude oscillatory shear (LAOS) deformation and 
extract the rheological response from the virial stress tensor (see Methods). 
\\

The 3-dimensional gel structures obtained from molecular simulation are characterized in terms of three topological invariants: the number of connected components $\beta_0$, of holes $\beta_1$, and of voids captured by $\beta_2$ (these so-called Betti numbers are combined to obtain the EC number) \cite{smith2021euler}. A void is an empty cavity within a shape that is surrounded on all sides by a solid boundary. To provide an intuitive understanding of the difference between voids and holes, we can think of an open cell versus closed cell gel or foam. An open cell structure contains many holes, whereas a closed cell structure contains many voids. To quantify the multi-scale structure of our soft gel simulations we perform filtrations using increasing particle diameter and quantify the topology of the resulting \v{C}ech complex (simplicial complex) through the EC. The filtration diameter is scaled such that $D=1$ represents the true diameter of the simulation particles. For example, a filtration radius $D = 2$ represents a Euclidean ball that is twice as wide as the original gel particle. We illustrate the filtration process on a gel simulation snapshot in Figure \ref{fig:3dfilt}, and we also visualize the topological changes for the full simulation, which contains $\sim$16,000 particles, as particle diameter is increased. We do this to illustrate how the various topological invariants: components, cycles, voids, could evolve during a filtration. 
\\

Figure \ref{fig:3dfilt} shows how the multi-scale structure of soft gels can be directly quantified through topology and filtrations. The topology of the gel goes through various topological phase transitions as the filtration diameter is increased. At $D = 1$ to $D = 2$ particle diameters we see a stable network topology. This stable topology reflects the bonded structure of the gel, and is often what is studied via network analysis tools. This network is dominated by cycles, and yields a negative Euler characteristic. As we pass to $D > 2$ particle diameters we see large topological phase transitions. Here we see a large increase in the number of cycles, an increasingly negative EC value, and the formation of an open cell structure. The measured structure now reflects the 3-dimensional structure (e.g., packing) of the gel and captures overlap between particles that are not necessarily directly bonded. Finally, at $D > 3.5$ particle diameters we see a transition from the open cell structure dominated by cycles to the closed cell structure that is dominated by voids driving the EC to a positive value. 
\\

\begin{figure}[!htp]
     \centering
     \begin{subfigure}[b]{0.49\textwidth}
         \centering
         \includegraphics[width=\textwidth]{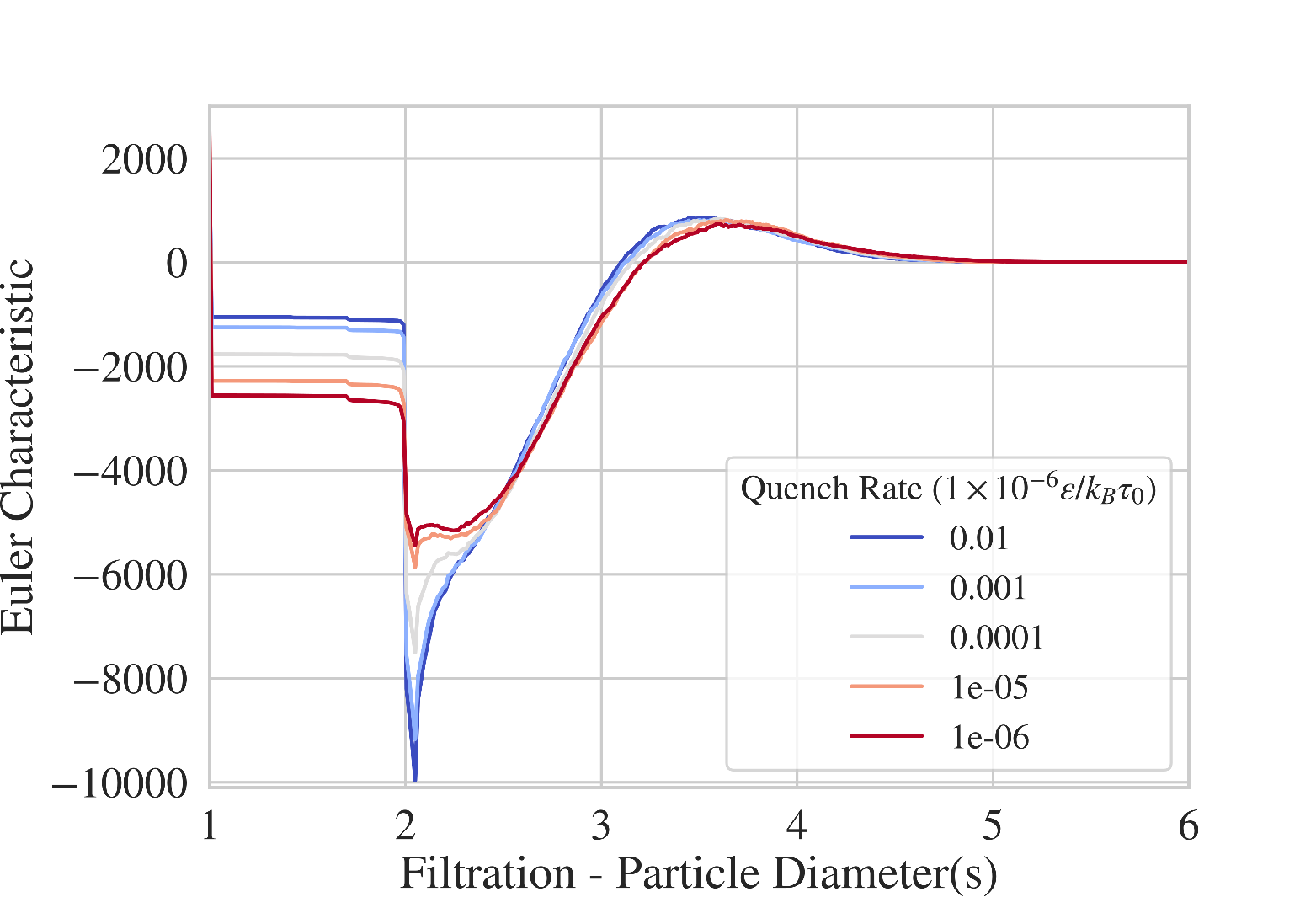}
         \caption{}
         \label{fig:q_change}
     \end{subfigure}\
     \hfill
     \begin{subfigure}[b]{0.49\textwidth}
         \centering
         \includegraphics[width=\textwidth]{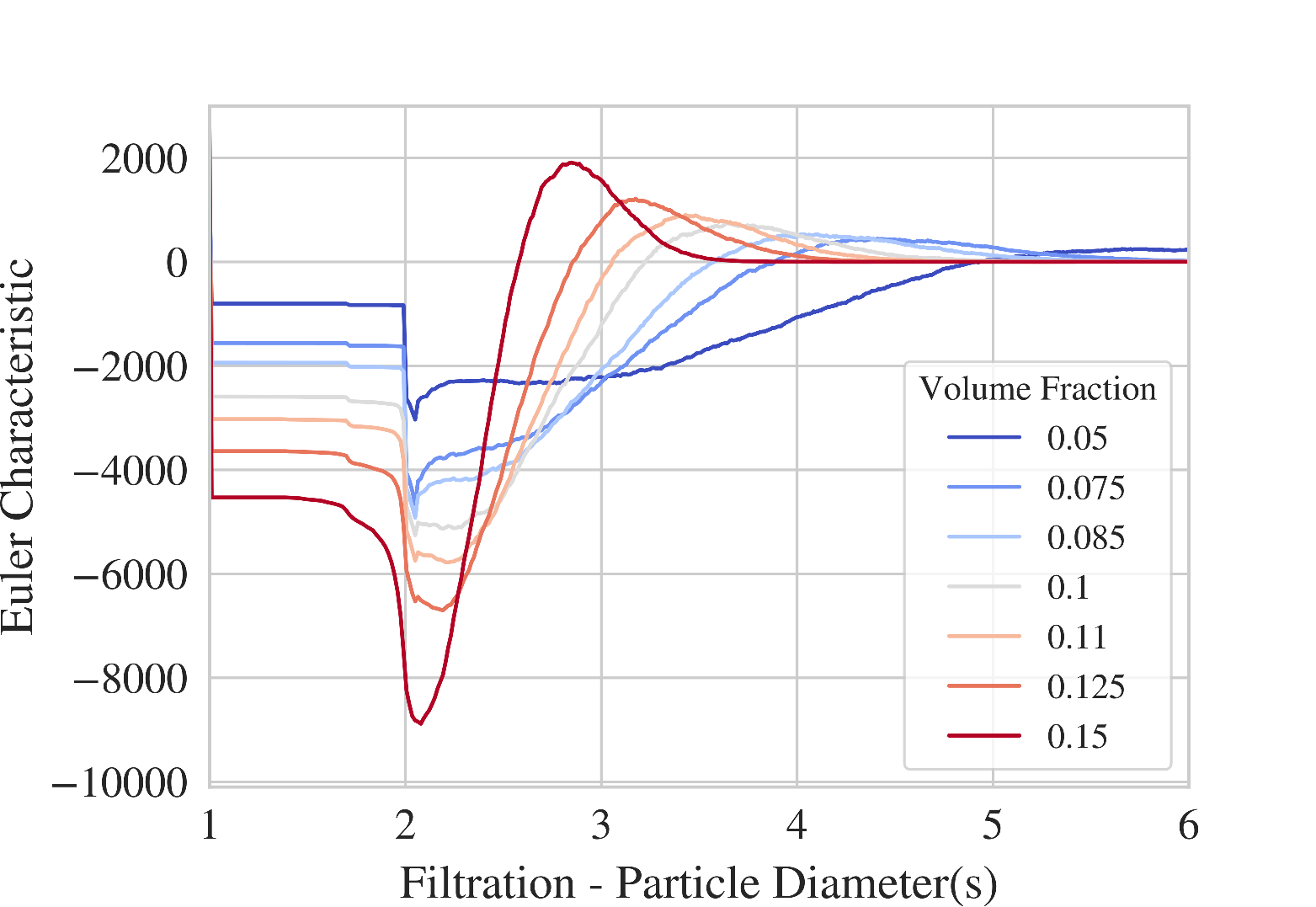}
         \caption{}
         \label{fig:v_change}
     \end{subfigure}
     	  \caption{(a) EC curves computed from gels at various quench rates with same volume fraction (0.1). (b) EC curves computed from gels at various volume fractions and same quench rate ($1\times 10^{-6} \epsilon/k_{B}\tau_{0}$, with $\epsilon$ and $\tau_{0}$ respectively the unit energy and time in the simulations). There is a continuous change in the topology/geometry of the gels as quench rate and volume fraction are changed. There are significant differences in the way the larger scale structure of the gel is changed when either volume fraction or quench rate is changed. This suggests that these parameters can be used to create bespoke gel structures.}
        \label{fig:var_vol}
\end{figure}

We leverage these topological characterizations to understand the relationship between soft gel structure and variance in the preparation parameters for the soft gel. In Figure \ref{fig:var_vol} we show a direct comparison of the EC curves for soft gels with varying volume fraction (for a constant quench rate at which the gel was formed in the simulations) and for soft gels formed with varying quench rates (at a constant volume fraction of $0.10$). We see that there is a smooth deformation of the EC curves with respect to changes in these two parameters, but the way in which these parameters change the gel structure is different. For the soft gels with varying quench rate shown in Figure \ref{fig:q_change}, we see that the changes in the soft gel are primarily impacting the local bonding structure ($D =1-2.3$ particle diameters) and we see minimal changes in the larger scale structure ($D>2.3$ particle diameters). This suggests that changing the quench rate may modify the local structure of the gel, but less so its mesoscopic 3-dimensional organization. Thus, quench rate could be used to tune/control the local organization of a soft gel without necessarily impacting its large-scale organization. In Figure \ref{fig:v_change} we see that the same is not true when volume fraction is changed; specifically, here we see that there is a large change in gel topology across multiple length scales. Interestingly, the impact on local bonding topology ($D = 1-2$ particle diameters) for increasing volume fraction follows a similar trend to decreasing quench rate, but diverges at larger scales ($D>2$ particle diameters). This suggests that a combination of quench rate and volume fraction tune/control could be used to create soft gels with target structures.  

Another important aspect in the application of TDA methods is computational scalability. The methods employed here are able to process gel simulations with 16,000 to 160,000 particles in a few minutes and on a single laptop computer. Other methods that attempt to capture similar information, such as minimal basis cycles of a graph, often run in polynomial time on the number of edges and nodes in a graph \cite{horton1987polynomial}. Furthermore, these graph-theoretic methods are focused at a single length scale; as such, attempting to perform these computations across various length scales would compound computational costs and affect overall scalability. Moreover these methods represent structure as a 2-dimensional object and thus might miss important information of how the network graph is embedded in 3-dimensional space. 

\subsection{Gel Dynamics - Shear Analysis} \label{topo_rheo}

\begin{figure}[!htp]
     \centering
     \begin{subfigure}[b]{0.49\textwidth}
         \centering
         \includegraphics[width=\textwidth]{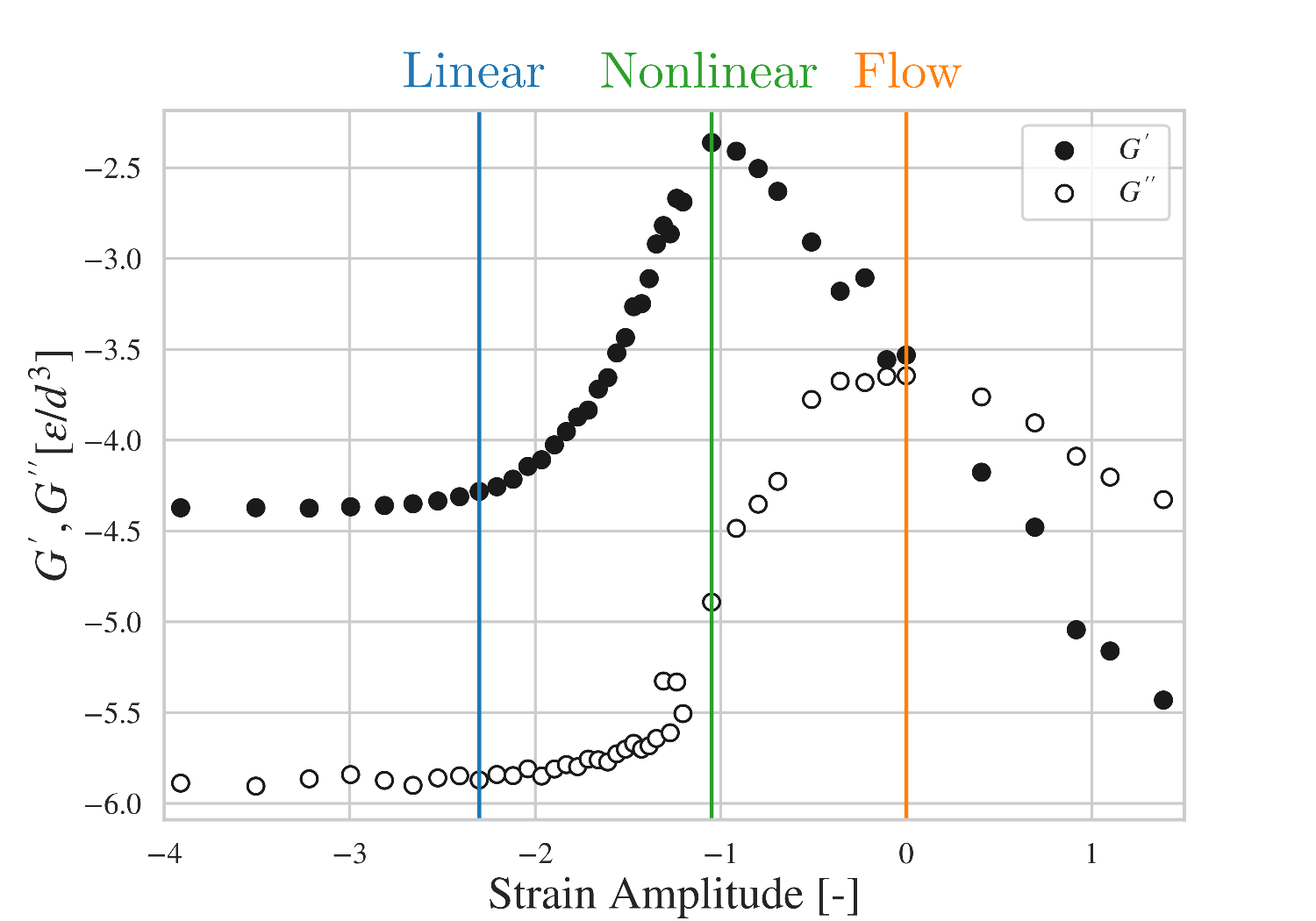}
         \caption{}
         \label{fig:dyn_ampsweep}
     \end{subfigure}
     \begin{subfigure}[b]{0.49\textwidth}
         \centering
         \includegraphics[width=\textwidth]{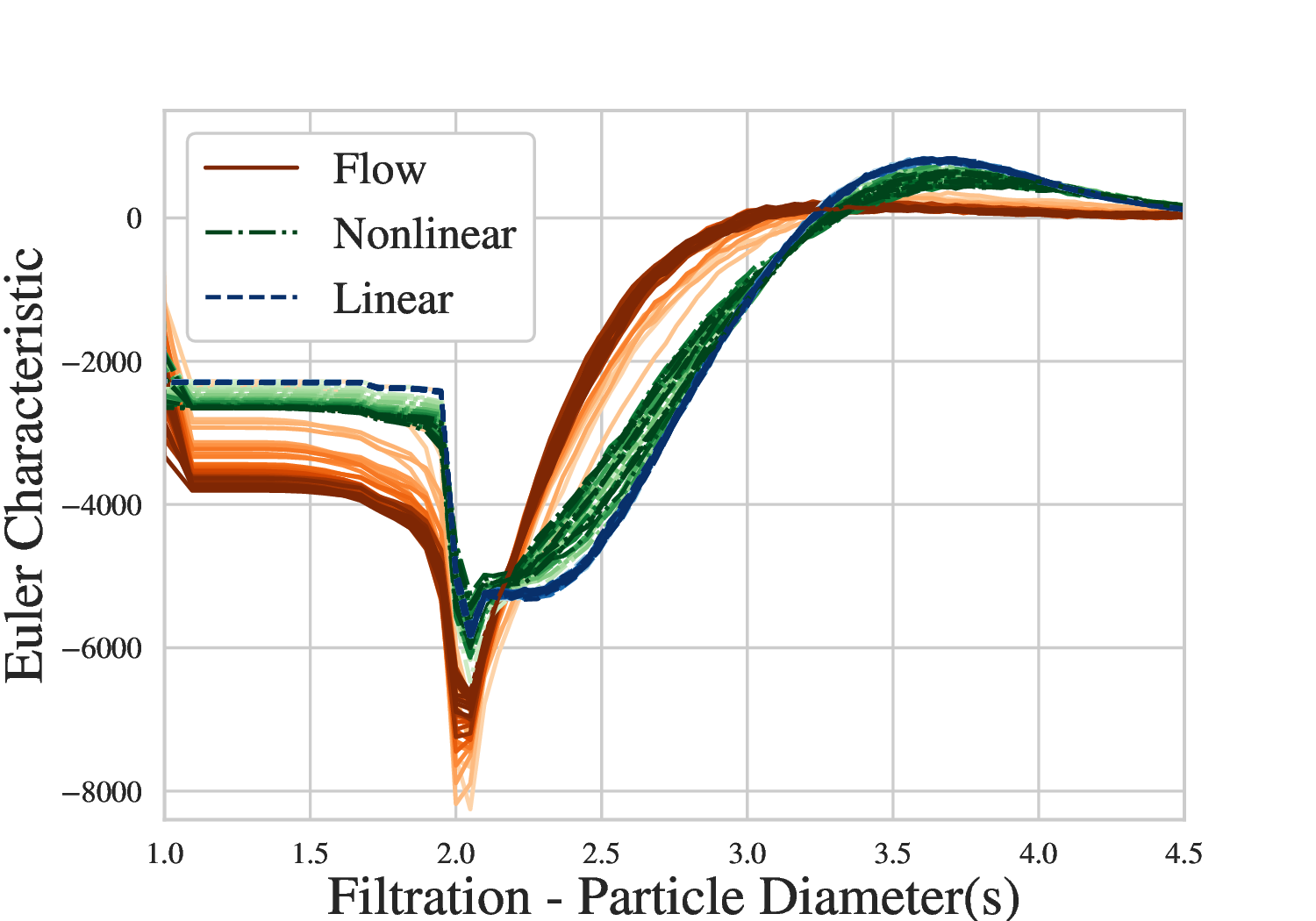}
         \caption{}
         \label{fig:dyn_ecs}
     \end{subfigure}\
     \begin{subfigure}[b]{0.49\textwidth}
         \centering
         \includegraphics[width=\textwidth]{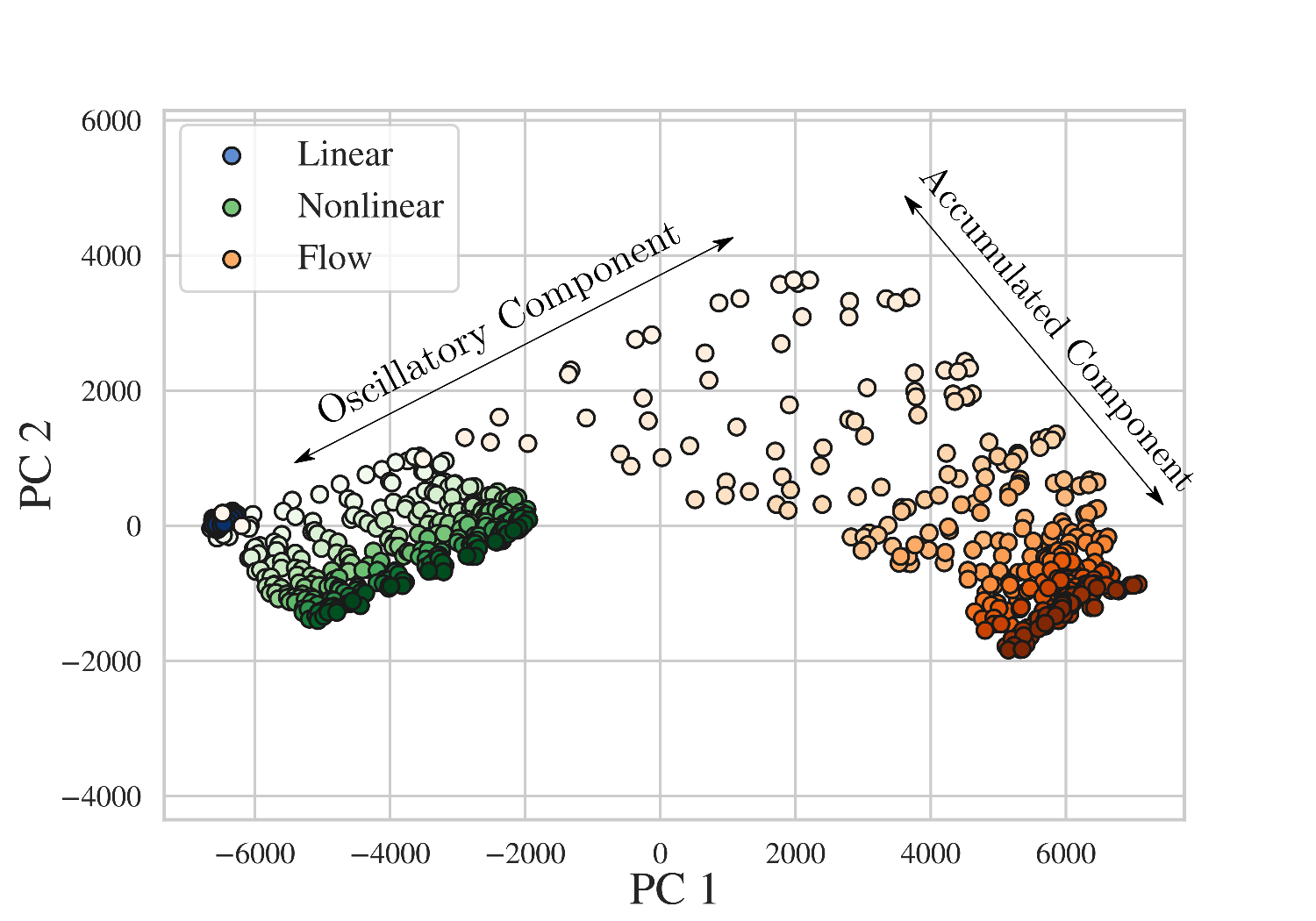}
         \caption{}
         \label{fig:pca_ecs}
     \end{subfigure}
     \begin{subfigure}[b]{0.49\textwidth}
         \centering
         \includegraphics[width=\textwidth]{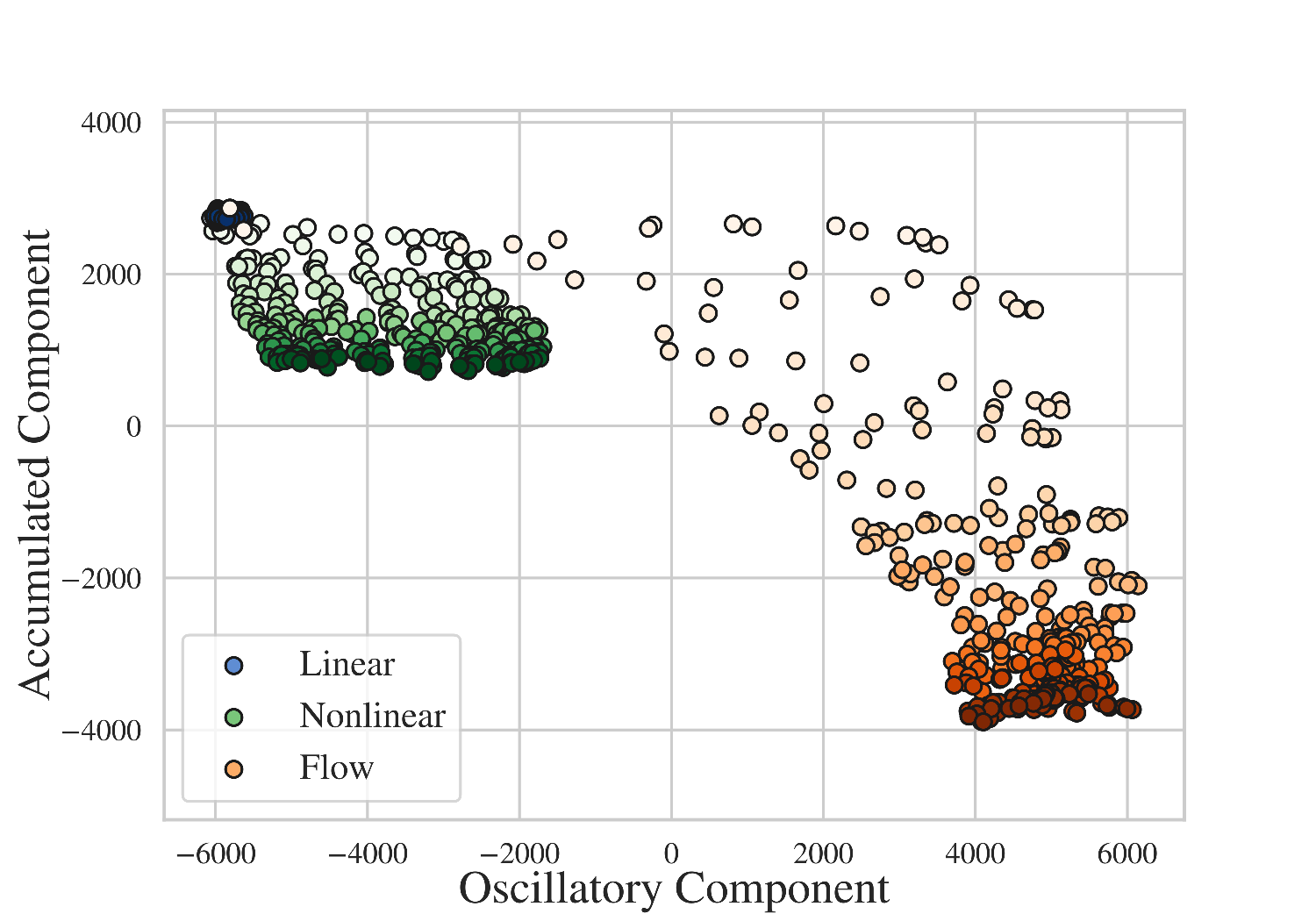}
         \caption{}
         \label{fig:pca_new}
     \end{subfigure}
        \caption{(a,b) EC curves computed at various temporal snapshots for a soft gel that is undergoing linear, nonlinear, and flow response to shear. The dynamic evolution is illustrated by difference in color, with the initial EC curves in the lightest color and the darkest color representing the final EC curve. We see that each simulation begins at the exact same EC curve but evolve very different as the shear amplitude is changed. (b) PCA applied to the EC curve data to create a low-dimensional representation of the dynamics of the gel during three different shear amplitudes. PCA captures $98\%$ of the variance in the data; we see two orthogonal dynamic modes that emerge, one associated with the fast oscillations of the gel induced by oscillatory shear and a slower mode capturing cumulative structure changes. (c) An optimal rigid rotation matrix is applied to the dynamic modes so that we may observe the cumulative and oscillatory components independently.}
        \label{fig:shear_pca}
\end{figure}

The ability to rapidly quantify the topology of these gel  structures allows us to perform high-throughput analysis of large temporal datasets, as those found in the analysis of gel rheology \cite{ruiz2019rheological, donley2022time}. Here, we explore the topological and geometric deformations of soft gels as they undergo oscillatory shear. The time-averaged amplitude sweep response of the gel studied here is shown in Figure \ref{fig:dyn_ampsweep}. Due to the density of information contained in the time-resolved rheological responses, we restrict our analyses to three key transition amplitudes  \cite{donley2022time}: the transitions (1) between the linear and non-linear elastic regimes, (2) the non-linear elastic and yielding regimes, and (3) the yielding and flow regimes. These transitions occur in this specific gel system at amplitudes of 0.1, 0.35, and 1.0 strain units,  respectively (see Figure \ref{fig:dyn_ampsweep}). At each of these amplitudes, we display time-resolved rheological and structural data extracted at a frequency of 32 points per period of oscillation.
\\

Figure \ref{fig:dyn_ecs} illustrates the dynamic evolution of EC curves for a soft gel undergoing various amplitudes of oscillatory shear that induce linear, nonlinear, and flow regime responses. The same initialized gel is used (volume fraction: 0.10, quench rate: $1\times10^{-6} \epsilon/k_{B} \tau_{0}$, using the simulations reduced units) but is exposed to different shear amplitudes. This figure also shows the initial EC curve (initial topology) in the lightest color and the darkest color representing the final EC curve (final topology). We see that each simulation begins at the exact same EC curve but evolves differently as the shear amplitude is changed, though the specific changes are difficult to discern. To more clearly visualize the changes in the EC curves as the system is sheared, we used PCA to project the EC curves to a low-dimensional space (see Methods section). Figure \ref{fig:pca_ecs} visualizes the projection of all EC curves onto the first two principal components (orthogonal dynamic modes) of the collective dataset. We found that the projection onto the first two orthogonal dynamic modes accounts for approximately 98 \% of the variance in the EC data.
 
Figure \ref{fig:pca_ecs} shows that PCA effectively captures the maximum amount of variation in the data with the minimum number of dynamic modes. This analysis reveals that there are two orthogonal dynamic behaviors captured in this low-dimensional representation: one that contains much of the fast oscillatory dynamics associated with the gel oscillation (we call this the oscillatory component) and one that contains a slow cumulative structural dynamic change within the soft gels, in particular for the nonlinear and flow regimes (we call this the cumulative component). To directly analyze these dynamic modes in the data, we can apply a rigid rotation matrix to the first two principal component so that we may observe the cumulative and oscillatory components independently. The resulting rotated visualization is found in Figure \ref{fig:pca_new}. Details around the rigid rotation and criterion for optimal rotation can be found in the Methods section. 
\\

\begin{figure}[!htp]
     \centering
     \begin{subfigure}[b]{0.4\textwidth}
         \centering
         \includegraphics[width=\textwidth]{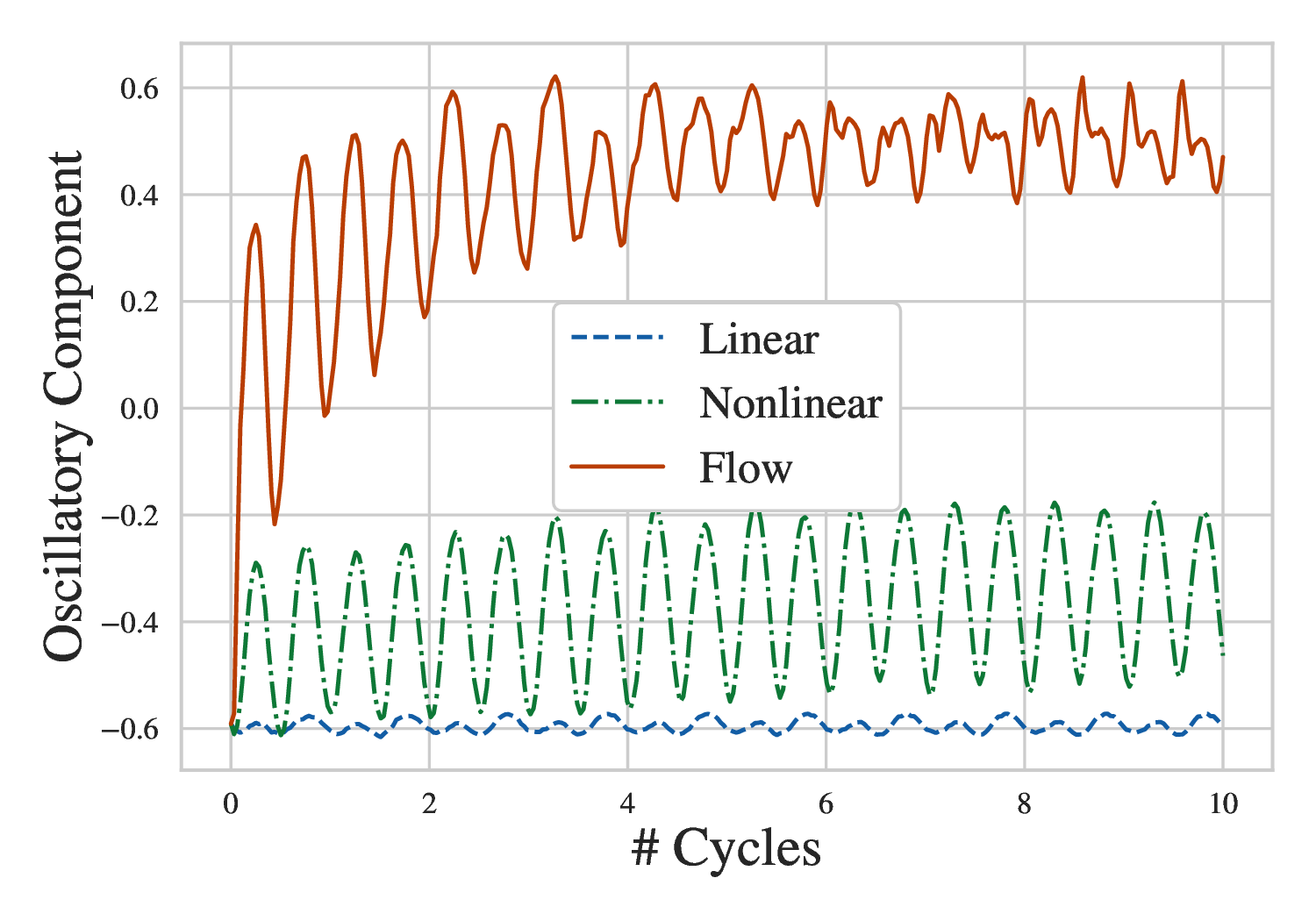}
         \caption{}
         \label{fig:osc_comp}
     \end{subfigure}\
   \hspace{0.2in}
     \begin{subfigure}[b]{0.4\textwidth}
         \centering
         \includegraphics[width=\textwidth]{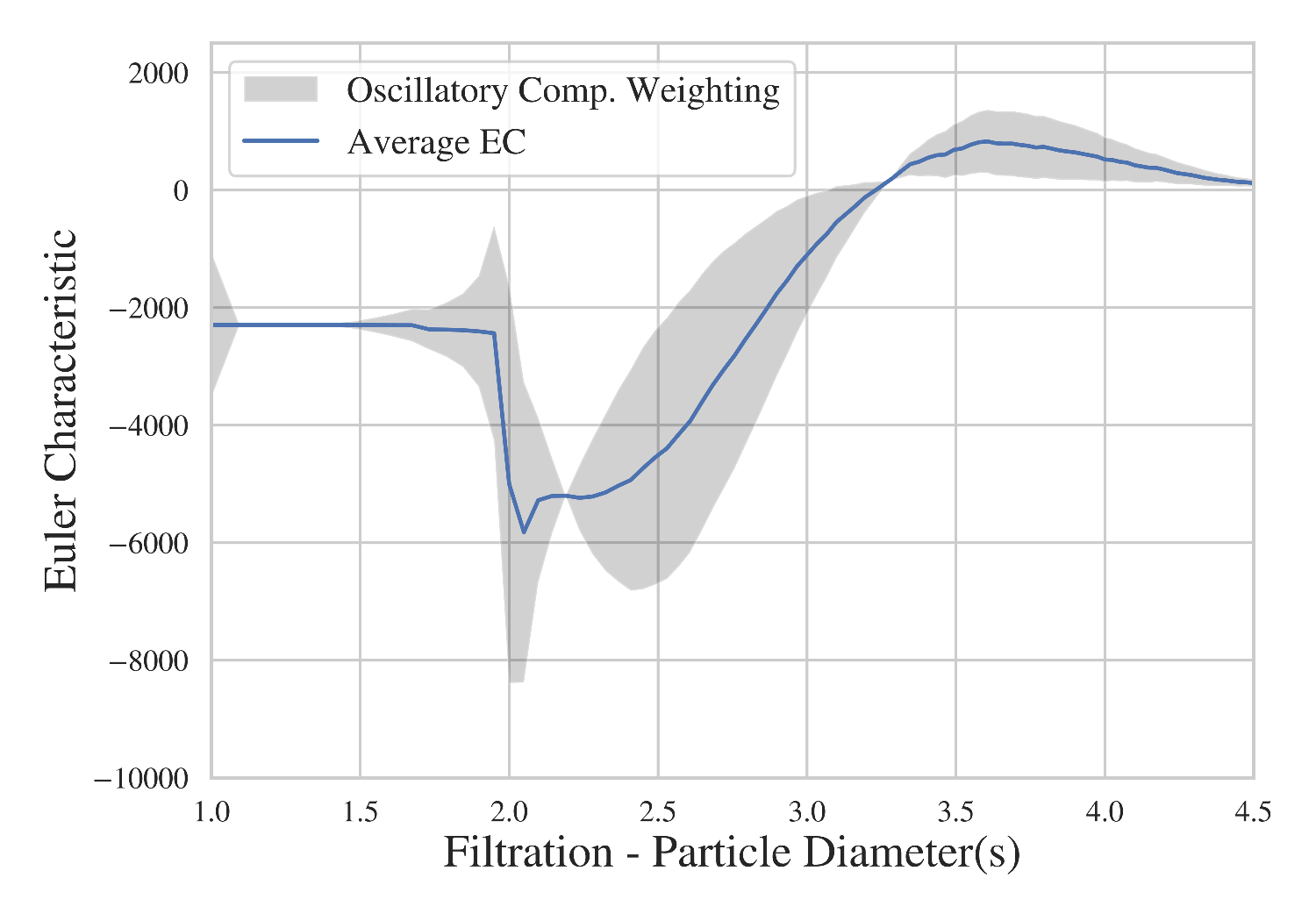}
         \caption{}
         \label{fig:osc_weight}
     \end{subfigure}
     \begin{subfigure}[b]{0.4\textwidth}
         \centering
         \includegraphics[width=\textwidth]{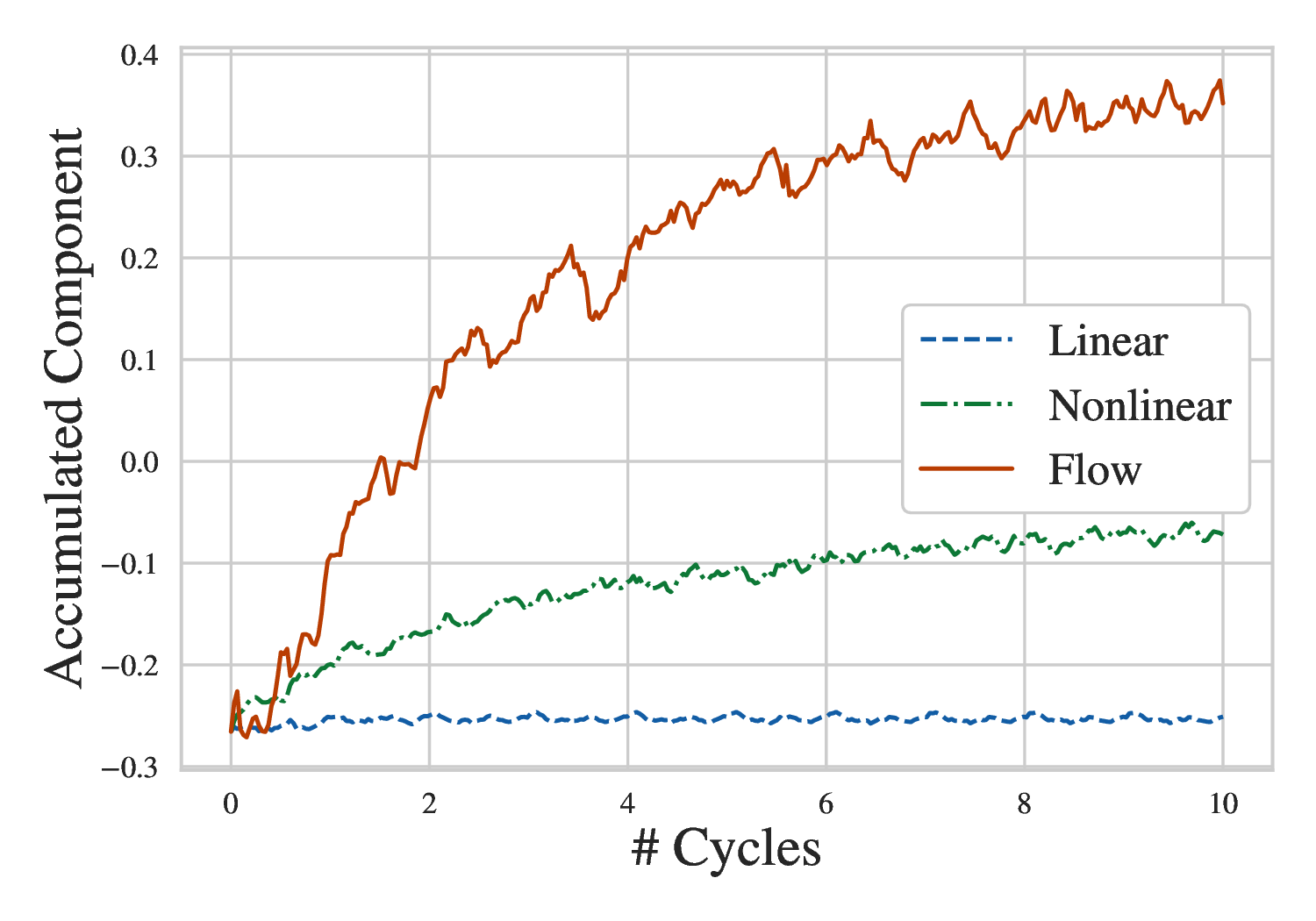}
         \caption{}
         \label{fig:temp_comp}
     \end{subfigure}\
   \hspace{0.2in}
     \begin{subfigure}[b]{0.4\textwidth}
         \centering
         \includegraphics[width=\textwidth]{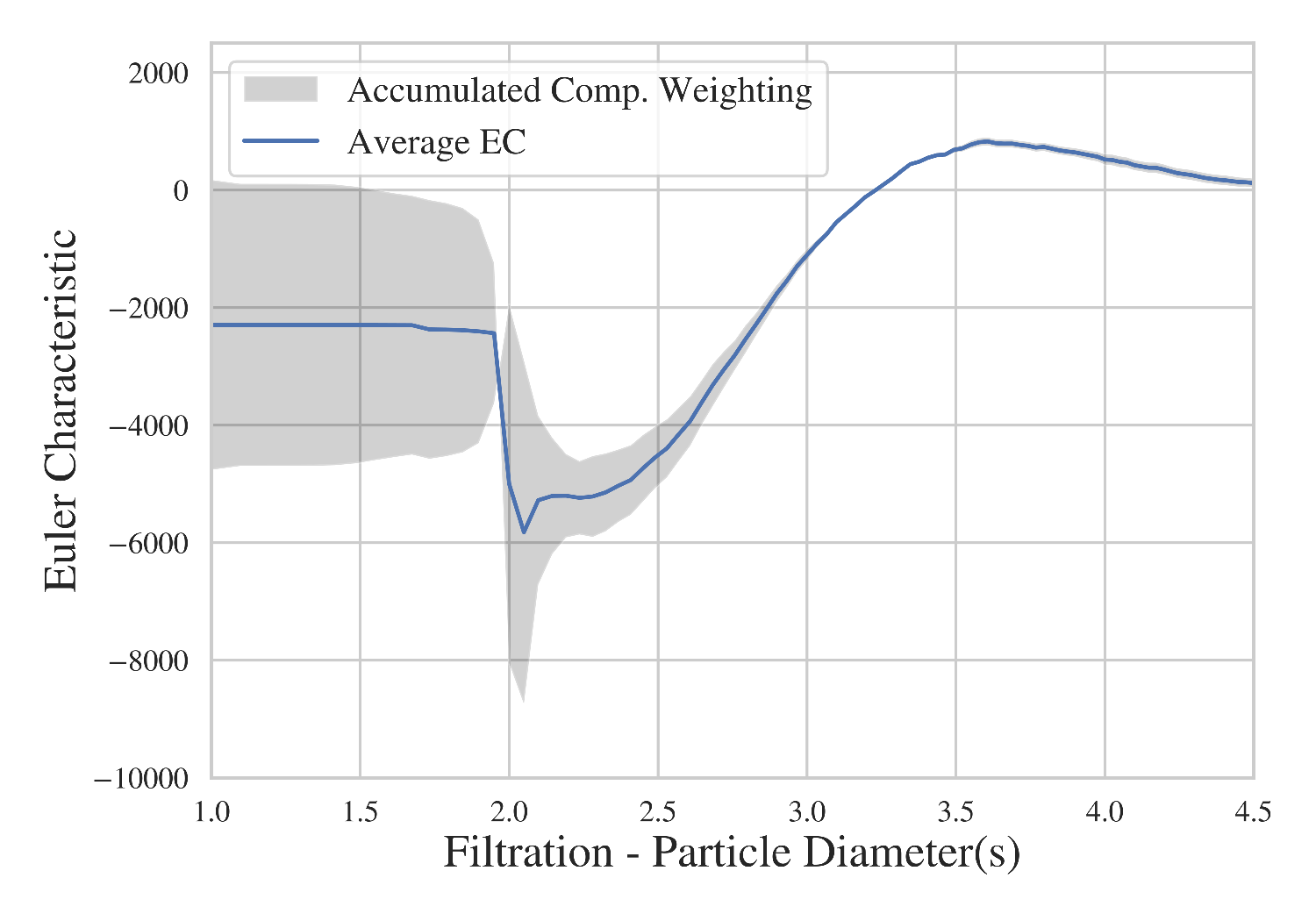}
         \caption{}
         \label{fig:temp_weight}
     \end{subfigure}
     \begin{subfigure}[b]{0.4\textwidth}
         \centering
         \includegraphics[width=\textwidth]{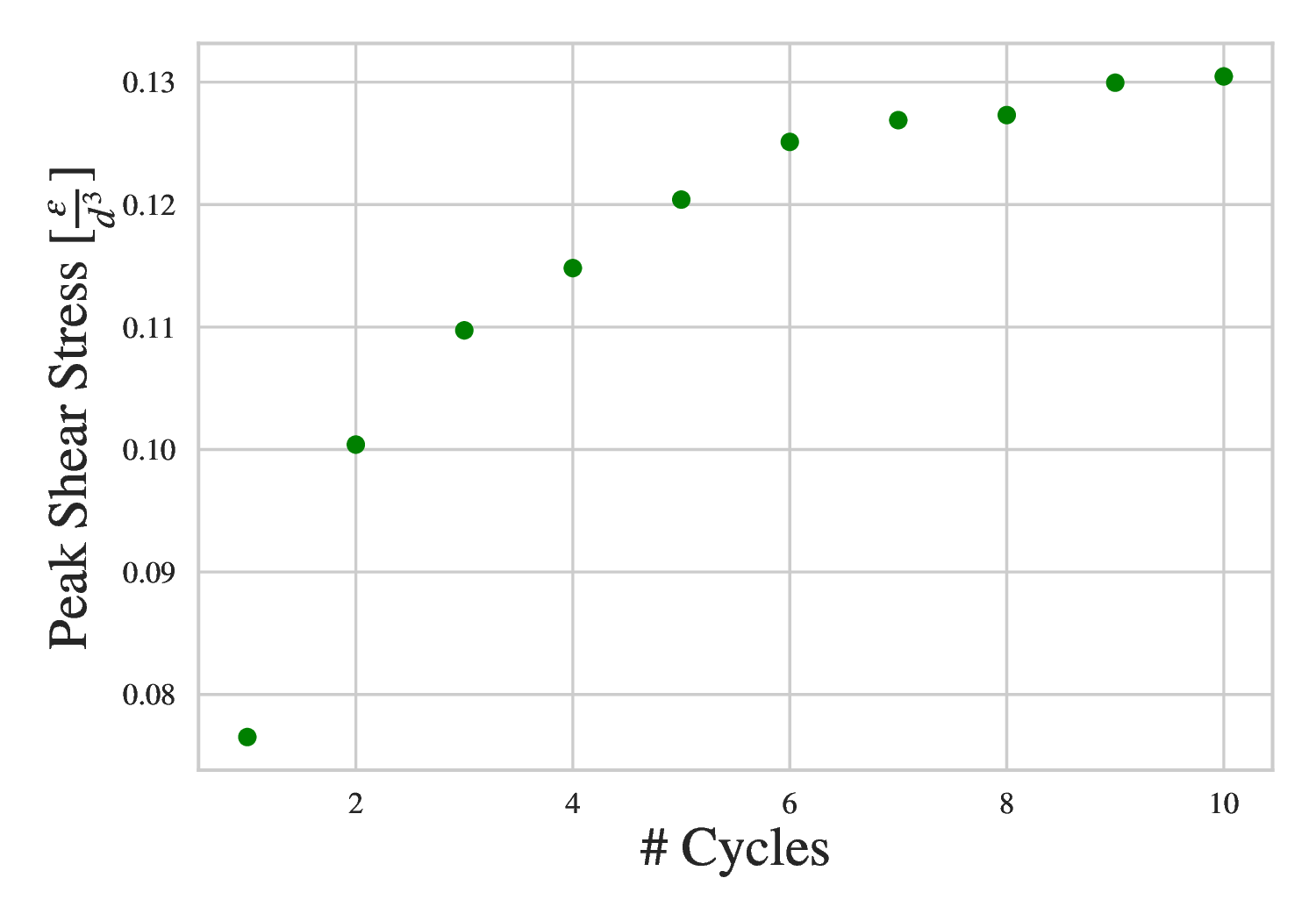}
         \caption{}
         \label{fig:shear_stress}
     \end{subfigure}\
     \hspace{0.2in}
     \begin{subfigure}[b]{0.4\textwidth}
         \centering
         \includegraphics[width=\textwidth]{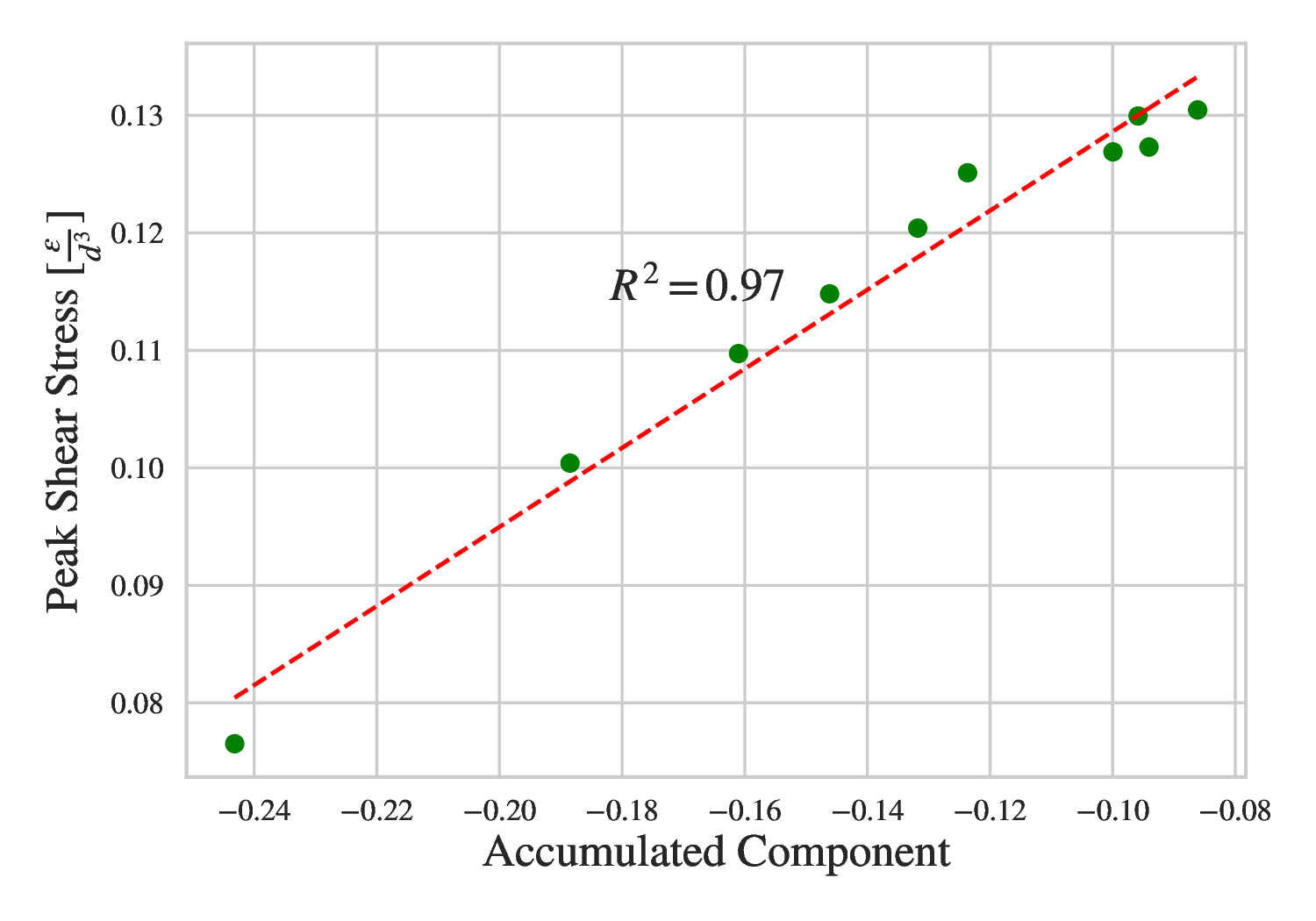}
         \caption{}
         \label{fig:shear_v_pc}
     \end{subfigure}
     \vspace{-0.1in}
        \caption{\small (a,c) Visualization of the oscillatory and cumulative component identified through PCA for gels undergoing oscillatory shear. Each component evolves over time, which illustrates the different dynamics expressed for linear, nonlinear, and flow regime responses. (b,d) Visualization of the oscillatory and cumulative component weightings for the EC curves. The area of the weightings (gray) represent higher emphasis placed on that particular portion of the EC. We see that the oscillatory component focuses on the mesoscale structure of the soft gel ($D > 2.3$ particle diameters), whereas the cumulative component focuses on the local bonding/network structure of the gel ($D < 2.3$ particle diameters). This suggests that these different scales of the material undergo different dynamics during oscillatory shear. (e,f) Demonstrates a clear correlation identified between the phenomena of shear stiffening of a gel responding nonlinearly, evidence by the increase in peak shear stress (e), and the dynamics of the cumulative component (c). We compute an $R^2 = 0.97$ which suggests that the cumulative component is an excellent predictor for the increase in peak shear stress and the phenomena of shear stiffening. \normalsize}
        \label{fig:shear_pcs}
\end{figure}

We transformed the high-dimensional set of EC curves into a couple of  simple scalar values that describe the dominating dynamic modes emerging during shearing. Figures \ref{fig:osc_comp} and \ref{fig:temp_comp} show how these values evolve during the three simulations. Because PCA is a linear method, we can identify what portions of the EC curve are associated with each dynamic mode by observing the weighting of the EC curve by the oscillatory and cumulative components shown in Figures \ref{fig:osc_weight} and \ref{fig:temp_weight}. Here, a larger gray area represents a higher weighting of that specific portion of the EC curve. This information reveals that the oscillatory component is capturing changes in the mesoscale structure of the soft gel ($D > 2.3$ particle diameters) while the cumulative component is capturing local changes in the soft gel network ($D < 2.3$ particle diameters).
\\

Figure \ref{fig:osc_comp} captures much of the oscillatory behavior that is induced by the oscillatory shearing of the gel. The linear response regime shows minimal deformation and all deformation seems to be elastic as there are no continuous offsets in the data. The same is true for the nonlinear response, but we see larger oscillations which are expected as the shear amplitude has increased. The flow response shows multiple interesting characteristics that differentiate it from the linear and nonlinear regimes. First, there is a large shift in topology in the first initial oscillations of the gel, this suggests that there is immediate deformation in the mesoscale topology of the gel during flow. Furthermore, we see that as shear continues the gel reaches a new \emph{topological} steady state in which the oscillations are no longer clearly defined, suggesting that there is a potential topological  transition occurring at the mesoscale in the material as it undergoes flow. Physically, these changes in the EC suggest a near immediate formation of larger cavities and holes within the gel without necessarily disrupting the local bonding structure. 
\\

Moving to the cumulative component shown in Figure \ref{fig:temp_comp}, we see that the linear regime shows almost no change over time. The nonlinear and flow regimes show similar dynamics but with a larger magnitude change for flow. We also note there is no dramatic offset in the flow response which is seen in the oscillatory component. This dynamic mode is associated with the local network/bonding structure of the soft gel, suggesting that this cumulative structural change is capturing the breaking and forming of bonds within the soft gel, which is supported by the weighting shown in Figure \ref{fig:temp_comp}. This mode is of particular interest because the structural changes are similar to those induced by reducing quench rate in gel preparation (Figure \ref{fig:q_change}), suggesting that this mode is capturing potential hardening of the gel. We explore the relationship between stiffening/hardening and the cumulative component in Figures \ref{fig:shear_stress} and \ref{fig:shear_v_pc}. Figure \ref{fig:shear_stress} shows the shear stress experienced by the soft gel at a shear amplitude of 0.350. This amplitude induces a nonlinear response in the gel and also stiffens the gel, this is evidenced by the increase in peak shear stress over time. Interestingly, we see a similar dynamic behavior in the cumulative component for the nonlinear response. We test whether there is a correlation between this dynamic mode and stiffening of the gel by comparing peak shear during oscillation and the corresponding value of the cumulative component at these time points, this is shown in Figure \ref{fig:shear_v_pc}. We see there is an obvious correlation between the variables ($R^2 = 0.97$), suggesting that the cumulative component could be used as a possible predictor of changes in peak shear stress. This also provides evidence that the mechanism for shear stiffening here is based on local bond reordering with minimal changes to the mesoscopic structure of the gel, similar to what is observed when quench rates are reduced. 
\\

Combining the analysis of these components also provides insight into why this same stiffening is not observed at the amplitude corresponding to the flow transition. The cumulative component at this amplitude shows a much greater change in the local bonding structure of the gel, which would suggest a greater stiffening of the gel given the relationship found in Figure \ref{fig:shear_v_pc} for the amplitude corresponding to the maximum non-linear elasticity. However, when comparing the oscillatory component between these amplitudes we see minimal non-elastic deformation in the mesoscale structure of the soft gel ($D > 2$ particle diameters) in the nonlinear regime, whereas in the flow regime there is a significant reorganization of the mesoscale structure of the soft gel, which may be weakening the gel overall. Similar evidence was found in \cite{colombo2014stress}, where significant bond breaking and forming was observed at high shear amplitudes (but the gel was becoming weaker). The authors hypothesized that there was a large scale reorganization of the gel, which is further supported by our topological analysis.
\\

\begin{figure}[!htp]
     \centering
     \includegraphics[width=\textwidth]{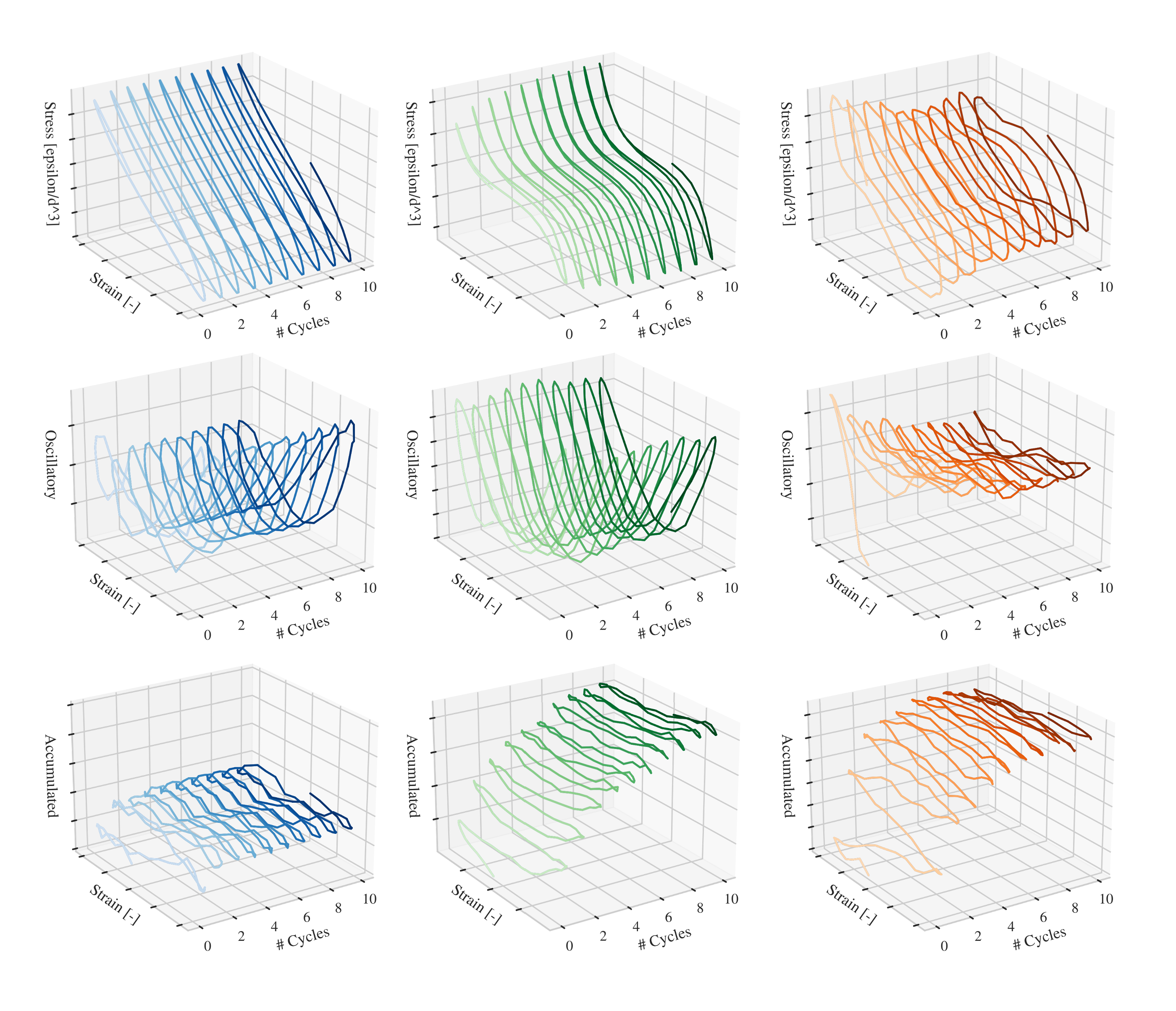}
     \caption{Comparison of the dynamics of 
 stress, the oscillatory component, and the cumulative component versus strain in the material for the linear (left column), nonlinear (middle column), and flow (right column) regimes. We observe similar changes in oscillatory dynamics as shear amplitude increases for both stress and the oscillatory component and a dramatic change in the oscillatory component during flow that is not experienced during the linear and nonlinear shear amplitudes. We also visualize the cumulative component showing how the total cumulative damage is smoothly increased over each oscillation of the gel in the nonlinear and flow regimes, while showing minimal change in the linear regime.}
     \label{fig:stress_strain_comp}
\end{figure}

We can further explore the physical connections by comparing the oscillatory and cumulative dynamics identified with PCA and the stress and strain undergone by the soft gel in Figure \ref{fig:stress_strain_comp}. Here, we qualitatively compare the evolution of stress vs. strain over each cycle and compare this to the dynamics of the oscillatory and cumulative component versus strain over each cycle. We observe strong similarities between the evolution of oscillatory component and stress when plotted against strain, illustrating similar changes in dynamics as shear amplitude increases and the dramatic change in the oscillatory component during flow that is not experienced during the linear and nonlinear shear amplitudes. This reinforces the physical connection between the topology of the soft gel and its rheological properties. We also see that the cumulative component smoothly increases over each oscillation during the nonlinear and flow regimes, while showing minimal change during the linear regime.  

\section{Discussion}

The topological and geometrical analysis of soft gels can be done effectively and efficiently through the use of topological data analysis (TDA). We have shown that the Euler characteristic (EC) of the \v{C}ech complex of the gel network, coupled with filtration operations, captures the multi-scale structure of soft gels that is missed by methods focusing independently on topology (e.g., bonding networks) or geometry (e.g., radial particle distributions in the physical space). We use EC curves as a concise summary of the topology and geometry of soft gels and use this information to understand the physical relationships between preparation parameters (e.g., quench rate, volume fraction). We find that there are clear continuous deformations in the structure of soft gels that correspond with changes in quench rate and volume fraction, but that the changes in the structure of the gel are not equivalent. In particular, we find that changing quench rate impacts the local organization of the gel and changes in volume fraction impact both local and global structure. This suggests that fine tuning of the quench rate and volume fraction could produce bespoke soft gels with state-of-the-art physical properties. It also indicates that applying our approach to gels formed through a range of kinetic processes can gain novel insight into the notoriously complex interplay between kinetics and gel structures. We explored the dynamics of soft gels under oscillatory shear and identified significant structural changes in soft gels that occur during linear, nonlinear, and flow responses. We identified a couple of dominant fast/slow dynamic modes within the sheared soft gels through the use of TDA and PCA. These modes represent dynamics of the gel at various length scales. We show that these modes can be directly correlated to physical phenomena such as shear stiffening or hardening, and provide insight into the multi-scale dynamics of complex nonlinear flow behaviors. Furthermore, these analysis methods are computationally efficient and scalable. We are able to perform computations on large simulations (in excess of 16,000 particles) in a few seconds on a single laptop computer. This is in stark contrast with other methods, such as minimum cycle basis algorithms, which scale nonlinearly in the number of particles/bonds (nodes/edges) found in a structure and only focus on a single length scale. The computational scalability and portability make the approach proposed here particularly interesting to interface data-mining and data-learning tools.
\\

This work primarily explores the topology of \v{C}ech complexes and filtrations to understand the structure of soft gels, but there are many other tools and methods from topology and geometry that can be incorporated into these analyses. The integration of topological concepts from knot theory, to start with, can help to understand the intertwining of soft gel structures and how the presence or absence of knots/links in the gel structure contribute to its physical characteristics \cite{horner2016knot}. The use of other methods from integral geometry, such as the full set of Minkowski functionals, can also help to understand better the full geometry of the gel structures as they undergo filtration. The Minkowski functionals (i.e., intrinsic volumes) represent orthogonal measures of intrinsic geometry and for 3-dimensional objects such as soft gels there are three independent measures: volume, surface area, and the EC \cite{mantz2008utilizing, hutter2003heterogeneity, koehl2023computing}. Future work could therefore explore how much information those other measures can gain. Finally,  the EC and more advanced topological characteristics such as those obtained from persistence homology, can be used to construct null-hypothesis and test for significant differences in soft gel structures \cite{dlotko2024topology}. 
\\

Overall, these methods can provide significant new insight into the behavior and potential design of new soft materials that involve complex and disordered network structures. Especially, there is often the need to identify where these specific topological structures exist within a given soft gel or material. While it is not possible to map directly from the EC curve back to the soft gel, methods such as the Hodge Laplacian can be used to identify the topological structures and understand what portions of a material are contributing to their physical behavior \cite{wei2022hodge}. The generalizability of the mathematical foundations for these methods makes them applicable to a wide range of data types, such as experimental images or videos. Recent advancements in 3-dimensional imaging of colloidal gels provides data that is directly amenable to the approaches proposed here and can be used to explore physical experiments, providing a route for direct comparison of molecular simulation and experiments \cite{dinsmore2002direct, dong2022direct, habdas2002video}. 

\section{Methods}

\subsection{Topological Data Analysis}

\begin{figure}[!htp]
    \begin{subfigure}{.3\textwidth}
      \centering
      \includegraphics[width=.06\linewidth]{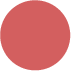}  
      \caption{}
      \label{fig:simplicial_complex}
    \end{subfigure}
    \begin{subfigure}{.3\textwidth}
      \centering
      \includegraphics[width=.4\linewidth]{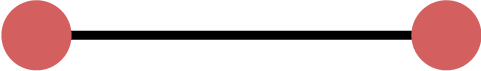}  
      \label{fig:Annulus}
      \caption{}
    \end{subfigure}
    \begin{subfigure}{.3\textwidth}
      \centering
      \includegraphics[width=.4\linewidth]{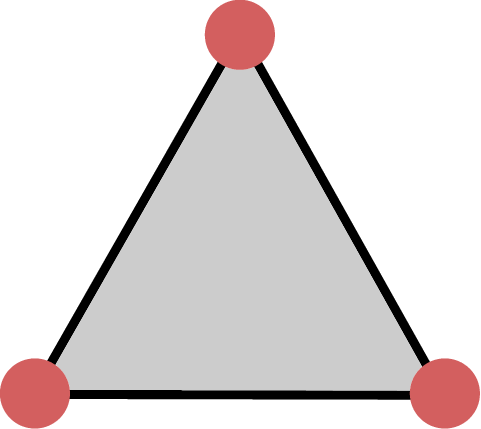}  
      \label{fig:Annulus}
      \caption{}
    \end{subfigure}
    \caption{Examples of $k$-dimensional simplexes. A simplex is a generalization of a triangle to higher (or lower) dimensions. (a) $0$-simplices are vertices (points), (b) $1$-simplices are edges, and (c) $2$-simplices are triangles.}
    \label{fig:simplex}
\end{figure}

\begin{figure}[!htp]
\centering
     \begin{subfigure}[b]{0.5\textwidth}
         \centering
         \includegraphics[width=\linewidth]{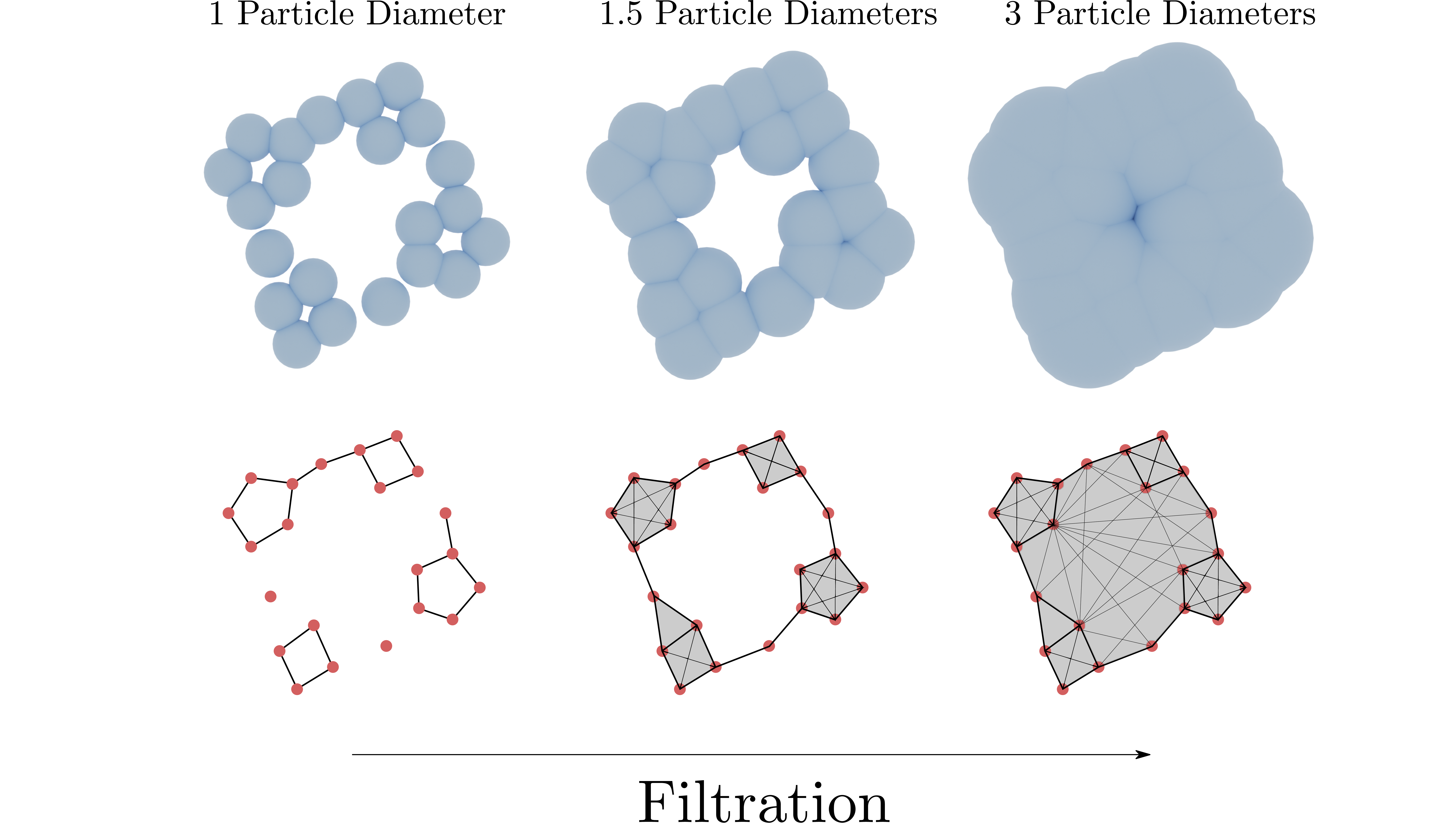}      
         \caption{}
         \label{fig:2dfiltcech}
     \end{subfigure}\
     \begin{subfigure}[b]{0.45\textwidth}
         \centering
         \includegraphics[width=\textwidth]{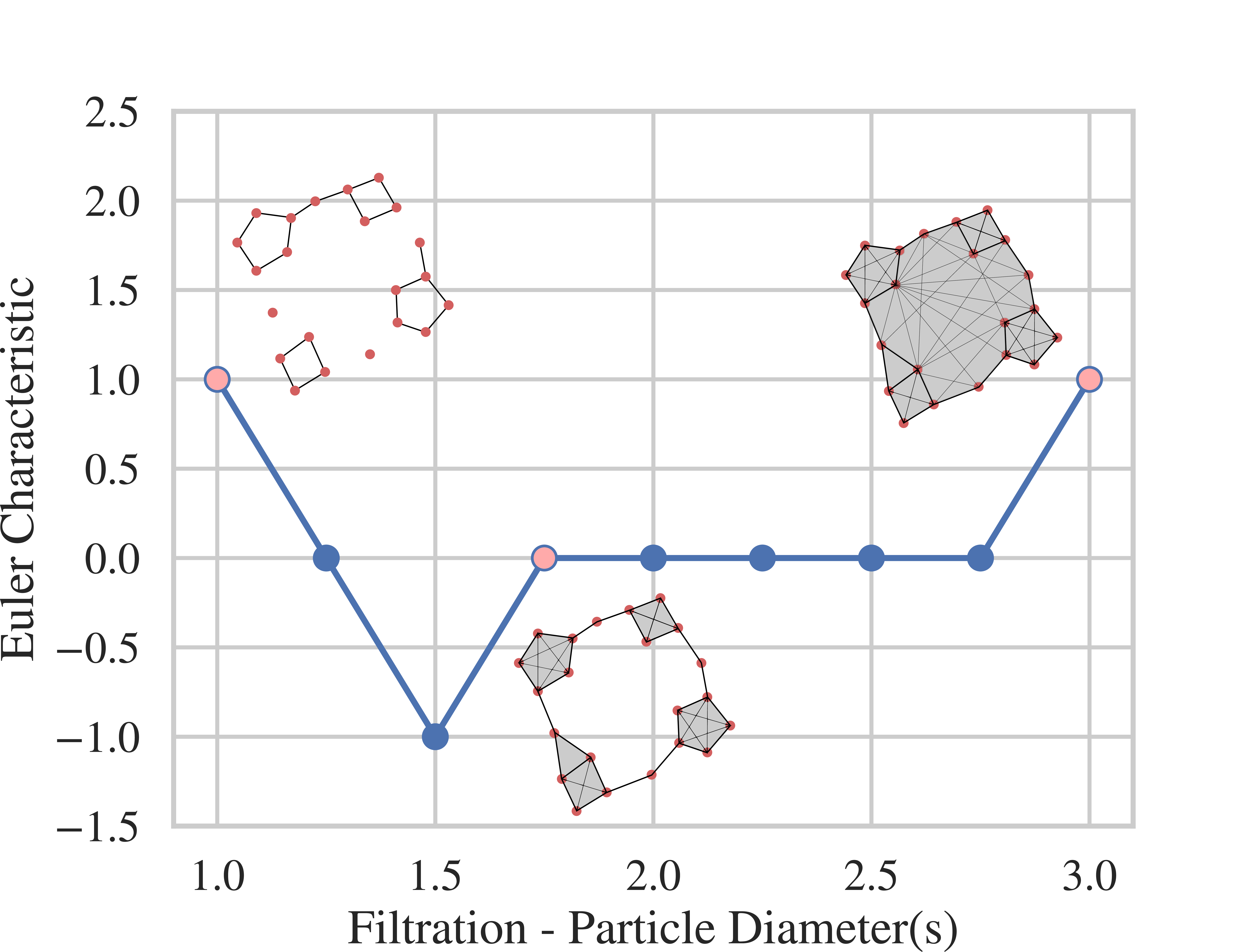}
         \caption{}
     \end{subfigure}
\caption{(a) Illustration of the changing topology of a 2-dimensional system as the diameter of the particles are increased. The overlap of the particles defines the formation of edges and higher-dimensional simplicies in the \v{C}ech complex (shown below). The \v{C}ech complex is a simplicial complex representation of the connectivity of the particles shown above. The simplicial complex representation allows us to compute and quantify the topology of the system. (b) The Euler characteristic (EC) curve computed from the filtration of structure shown in (a). We measure the EC $\chi(D)$ of the \v{C}ech complex at increasing particle diameters $D_i \in \mathbb{R}$. The ordered pairing of these values $\{\chi(D_i),D_i\}$ is the EC curve. The EC curve summarizes the topological and geometrical changes in the structure at various scales.}
\label{fig:2dfilt}
\end{figure}

Given a set of points $v \in \mathbb{R}^m$, in our case the positions of the center of mass of the particles that constitute a gel, we construct a \emph{simplicial complex} built from $k$-\textit{simplices} of varying dimension $k$. A $k$-\textit{simplex} is a convex set spanned by $k$+1 affinely independent points, denoted as:
\begin{align}
    \sigma = \{v_1,v_2,...,v_n,v_{k+1}\}
\end{align}
Simplices of varying dimension are illustrated in Figure \ref{fig:simplex}. A \textit{simplicial complex} (denoted as $\mathcal{K}$) is obtained by connecting simplicies of varying dimension and can be used to describe the topology of complex shapes. In Figure \ref{fig:2dfilt}, we use a simple 2-dimensional soft gel structure to illustrate the application of simplicial complexes to soft gel simulations. Here, the soft gel particles are represented as points in 2-dimensional space $x_i \in \mathbb{R}^2$, with a diameter $D = 1$. When a couple of particles overlap we add an edge (1-simplex) to our simplicial complex; three a triangle (2-simplex), and so on. The resulting simplicial complex is known as the \emph{nerve} of the overlapping particles, and more specifically the \v{C}ech complex (see next section for more details). Simplicial complexes defined in this way are exact representations of the topology of the particle system at a given particle diameter (same number of connected component, cycles, voids, etc.) while being able to be encoded and measured using simple algebraic operations. In Figure \ref{fig:2dfiltcech}, we show how at a $D=1$ particle diameter, many of the particles overlap with their immediate neighbors. Here, we begin to form edges or $1$-dimensional simplices that begin to capture a graph structure in our data which results in four small holes and 5 different connected components, reflecting a common network representation of a soft gel. However, we can see that these holes and connected components are not randomly distributed in space, there is a larger cyclic structure that is not captured when we consider connectivity at a single particle diameter. As we move from $D=1$ particle diameters to $D=1.5$ particle diameters we see instances where $3$ particles begin to overlap, which yields the formation of $2$-dimensional simplices (e.g., filled triangles). As we continue to increase $D$ we see increasing levels of overlap and our resulting simplicial complex is increasingly connected, revealing the larger cyclic structure captured within the \v{C}ech complex. Eventually we will reach a point where every particle is overlapped and we end up with a fully connected convex hull of our data, shown at $D = 3$ particle diameters. Through the \v{C}ech complex we gain an understanding of the geometry and topology of the soft gel at varying length scales (diameters). 

\subsubsection{\v{C}ech Complex}

To construct a \v{C}ech complex from our soft gel molecular simulations, we place at the center of each point (i.e., particle) $x_i \in \mathbb{R}^n$ an $n$-dimensional Euclidean ball $B(x_i,D) := \{y \in \mathbb{R}^n : ||x_i - y||_2 < D/2\}$ of diameter $D \in \mathbb{R}$. The set of these Euclidean balls is known as a \emph{cover} $\mathcal{U} := \cup\{B_{i}\}_{i\in I}$ of the points $x_i$, where $I \in \mathbb{Z}$ represents the total number of particles in our simulation. As we increase the diameter $D$ of the Euclidean balls $B(x_i,D)$, we will obtain some level of overlap between balls. This overlap is what defines the connectivity, and subsequently the geometry and topology of our data. Thus, from the cover of our points $\mathcal{U}$ we obtain a shape that describes the structure of our soft gel at a length scale defined by our ball radius $D$ (see Figure \ref{fig:2dfilt}). From the cover of balls, we can construct a \emph{nerve} which is a simplicial complex that represents exactly the topology of the overlapping balls. 

The nerve of collection $\mathcal{U} := \cup\{B_{i\in I}\}$ is the simplicial complex with vertices $I$ and $k$-simplices built from $\{i_0,i_1,...,i_k\}$ if and only if $B_{i_0} \cap B_{i_1} \cap ... \cap B_{i_k} \neq \{\varnothing\}$. The nerve of the cover $\mathcal{U}$ is known as the \v{C}ech complex that is defined by the particle positions and selected ball diameter. We then quantify the topology of the \v{C}ech complex through the Euler characteristic.

\subsubsection{The Euler Characteristic}

The Euler characteristic (EC), introduced by Leonard Euler in 1758, is a scalar value that quantifies the \emph{topological invariants} of a shape. A topological invariant is a characteristic of a shape that is unaffected by continuous deformation of the shape (e.g., stretching, bending), but is affected by discontinuous deformations (e.g., cutting, glueing). In our work, we are focused on 3-dimensional objects, which means that we are interested in three particular topological invariants: connected components (1-dimensional), holes (2-dimensional), and voids (3-dimensional). The total number of unique $i$-dimensional invariants in a shape are referred to as the Betti numbers $\beta_i$ where $i$ represents the dimension of the invariant being counted. The EC $\chi$ is defined as the alternating sum of the Betti numbers:
\begin{align}
    \chi = \sum_{i=1}^n (-1)^i \beta_i
\end{align}
Computationally, there are many ways to quantify the Betti numbers and the overall EC of a given \v{C}ech complex nerve \cite{smith2021topological, edelsbrunner2008persistent, edelsbrunner2022computational}. These methods range from the use of simplicial algebra to discrete Hodge theory \cite{knill2017one}. One of the simplest methods (we use this in our work) is an extension of Euler's original formula for polyhedra to simplicies given by \cite{richeson2019euler}:
\begin{align}
    \chi = \sum_{i=1}^n (-1)^i k_i
\end{align}
where $k_i$ represents the number of $k$-dimensional simplicies in the \v{C}ech complex. 

\subsubsection{Filtration and Euler Characteristic Curves}

The EC is able to effectively quantify the topology of a shape, but it does not account for all possible particle diameters and is thus sensitive to the diameter selected. We can account for this issue by applying a \emph{filtration} to our data where we measure the EC at multiple, increasing particle diameters. For example, 
through the analysis of the simplified 2-dimensional system in Figure \ref{fig:2dfilt}, we notice how the topology of the resulting \v{C}ech complex is sensitive to the selection of particle diameter $D$. This change in topology with respect to diameter provides us an understanding of the multi-scale nature of the system. For example, we see that at a radius $D = 1$ particle diameters, we capture multiple small cycles in the structure, while at a larger radius $D = 1.5$ particle diameters we see that there is a much larger cycle captured. For a set of increasing diameters $D_1 < D_2 < ... < D_l$, we can compute the EC $\chi(D)$ of the resulting \v{C}ech complex at each diameter. From this, we can create an ordered pairing of these values to construct an Euler characteristic curve. An example curve for our simple 2-dimensional gel is found in Figure \ref{fig:2dfilt}. These curves summarize the evolution of the soft gel topology as we vary the diameter of the particles (moving from local to global structures). We can see in Figure \ref{fig:2dfilt} that we begin our filtration at $D = 1$ particle diameter which captures a series of 4 small cycles and a total of 5 connected components giving $\chi(1) = 5 - 1$, this structure reflects what might be analyzed in a network representation of the system. We then reach a threshold at $D = 1.5$ where we see that the smaller cycles in our graph are collapsed and we retain a single large cycle, $\chi(1.5) = 1 - 1$. Eventually the \v{C}ech complex becomes fully connected and we end with a single connected component at $D = 3$ particle diameters, $\chi(3) = 1-0$. We can see from Figure \ref{fig:2dfilt} that the EC curve quantifies and summarizes these various topological changes. Furthermore, the EC curve can be represented directly as a vector, which can be integrated into common data analysis methods such as PCA and for use in quantifying relationships between the structure of the system and its physical behavior. 

\subsection{Principal Component Analysis}

To apply PCA we construct vectors $X_t \in \mathbb{R}^n$ of EC values $\chi(D) \in \mathbb{Z}$ of the \v{C}ech complex measured at $n$ diameters $D \in \mathbb{R}$ for each simulation snapshot time point $t \in \mathbb{Z}$. This vector $X_t := [\chi(D_1), \chi(D_2), ..., \chi(D_n)]$ represents the EC values measured at increasing diameter $D_1 < D_2 < ... < D_n$. From this we can construct a matrix $\mathbb{M}$ by stacking each of our $t \in \mathbb{Z}$ simulation snapshot EC curves: $\mathbf{M} := [X_1^T, X_2^T, ... X_t^T]^T \in \mathbb{R}^{t \times n}$. We construct a matrix $\mathbf{M}_{i}$ for each shear amplitude $i \in {0.100,0.350,1.00}$. In order to ensure the PCA projections are consistent across all simulations we perform PCA on all $\mathbf{M}_{i}$ simultaneously so that the derived principal components are the same for each simulation. Figure \ref{fig:pca_ecs} visualizes the projection of all simulation snapshot EC curves onto the first two principal components of the collective $\mathbf{M}_{i}$.

\begin{figure}[!htp]
     \centering
     \begin{subfigure}[b]{0.49\textwidth}
         \centering
         \includegraphics[width=\textwidth]{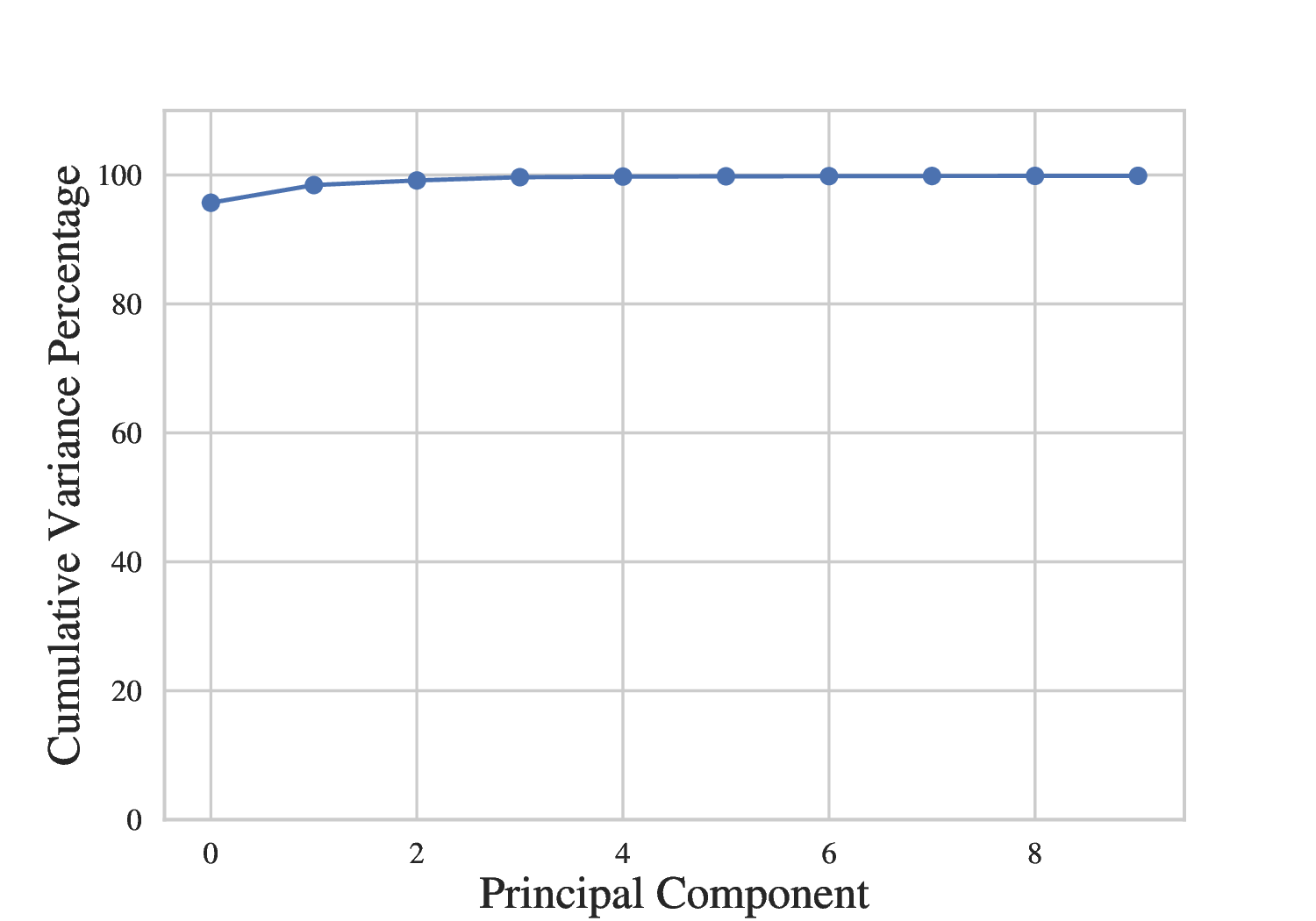}
         \caption{}
         \label{fig:pca_var}
     \end{subfigure}\
     \hfill
     \begin{subfigure}[b]{0.49\textwidth}
         \centering
         \includegraphics[width=\textwidth]{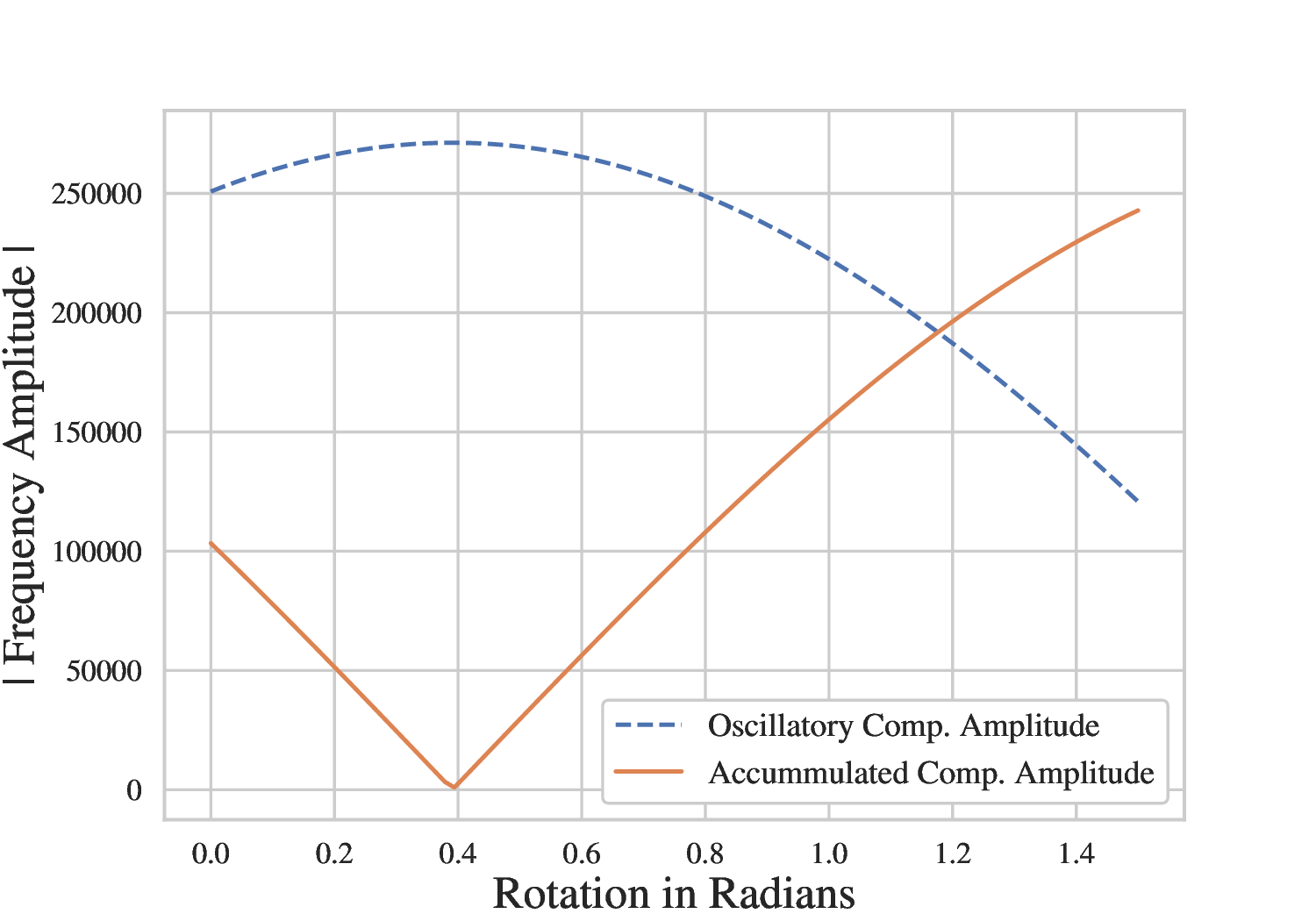}
         \caption{}
         \label{fig:pca_rot}
     \end{subfigure}
     	  \caption{(a) Cumulative explained variance percentage captured with consecutive leading principal components. It is shown that the first two principal components explain over 98$\%$ of the total variance within the dataset (b) Illustration of the absolute amplitude of oscillation contained within the first and second component as they are rotated. We see that there is an obvious minimum amplitude in the cumulative component and maximum amplitude in the oscillatory component at a rotation of 0.40 radians which is used in our analysis.}
        \label{fig:pca_just}
\end{figure}

The principal component analysis conducted provides an effective and interpretable dimensionality reduction for the EC curve data measured from the soft gel simulations. Here, we report the total explained variance captured in the 10 leading principal components and justify the optimal rotation of the first two principal components to separate the oscillatory and cumulative components. Figure \ref{fig:pca_var} shows the cumulative variance percentage contained in the leading 10 principal components. We see that the first two components capture over 98$\%$ of the total variance in the dataset supporting our selection of the first two principal components for our analysis. Figure \ref{fig:pca_rot} illustrates the optimal selection of rotation, in radians, for the data projected on the first two principal components in order to identify the oscillatory and cumulative components which are used in our analysis. Given data projected onto the first two principal components $\mathbf{v} = [v_1,v_2] \in \mathbb{R}^{2\times320}$, we apply a rotation matrix $\mathbf{R}(\theta)$ given as:

\[
\mathbf{R}(\theta) := \begin{bmatrix}
\cos(\theta) & -\sin(\theta) \\
\sin(\theta) & \cos(\theta)
\end{bmatrix}
\]

where $\theta$ represents the rotation angle in radians. We can identify the optimal rotation $\hat{\theta}$ by finding the $\theta$ value that maximizes the amplitude of oscillation in the oscillatory component and minimizes the oscillatory amplitude in the cumulative component. We measure the oscillatory amplitude through a simple Fourier transform. Figure \ref{fig:pca_rot} shows how these two values evolve as $\theta$ is increased from 0, showing an optimal value $\hat{\theta} = 0.4$. Thus, we obtain and optimally rotated principal component projection as $\tilde{\mathbf{v}} = \mathbf{R}(.4)\mathbf{v}$.

\subsection{Numerical Gel Model and Simulation}

We use an established numerical model for particulate gels \cite{bantawa21,bantawa23,bouzid2018network,colombo2014stress,donley2022time} consisting of self-assembling spherical particles of diameter $d$ that interact via a short-range attraction, $U_2$ and a three body term, $U_3$ which limits the bond angles and introduces a bending stiffness. This three-body term is meant to model hindrance of the relative particle motion upon aggregation, arising from sources such as particle surface roughness or the irregularly shaped aggregates. The Molecular Dynamics (MD) simulations are implemented in a system of $N$ particles with position vectors \{$\mathbf{r}_1,..., \mathbf{r}_N$\} and interacting with the potential energy:
\begin{equation}
U\left(\mathbf{r}_1,..., \mathbf{r}_N\right)=\epsilon \left[ \sum_{i>j}U_2\left(\frac{\textbf{ r}_{ij}}{d}\right)+\sum_i\sum_{j>k}^{j,k\neq i}U_3\left(\frac{\textbf{r}_{ij}}{d},\frac{\textbf{r}_{ik}}{d}\right)\right]
\label{Potential}
\end{equation}
where $\textbf{r}_{ij}=\textbf{r}_j-\textbf{r}_i$, $\epsilon$ is the depth of the attractive well $U_2$ and sets the energy scale, and $d$ is the particle diameter, representing the unit length scale. In typical (colloidal) systems, $d$ corresponds approximately to the range $d \simeq10$ to $100$~nm and $\epsilon \simeq 10$ to $100$~$k_{B}T$, where $k_B$ is the Boltzman constant and T, typically room temperature. The functional forms of the two-body ($U_2$) and three-body ($U_3$) terms in Eq. \eqref{Potential} are detailed elsewhere \cite{bantawa21,bouzid2018network,colombo2014stress,donley2022time}.

\subsection{Gel Preparation}

The preparation of the particulate gel configurations studied here has been detailed in prior works \cite{bouzid2017elastically,bouzid2018network,donley2022time}; here we cover the basics of the process. The preparation protocol consists of two parts. 

In a cubic simulation box of size $L$, we start from particles initially equilibrated at $k_BT/\epsilon=0.5$ and brought into a spontaneously self-assembled network at $k_{B}T/\epsilon =0.05$ through NVT (using a Nose-Hoover thermostat). The network is composed of strands (particles that have coordination number $z=2$) connected by branching points ($z=3$) and the system is further equilibrated for additional $2\cdot10^4$ MD steps. 

In the second part of the gel preparation, a damped dynamics is used to bring the configuration obtained as just described to a local minimum that more likely corresponds to a mechanically stable state. This is achieved by withdrawing the kinetic energy of the system to $\sim 10^{-10}$ of its initial value with an overdamped dissipative dynamics:
\begin{equation}
m\frac{d^2\textbf{r}_i}{dt^2}=-\nabla_{\textbf{r}_i}U-\zeta\frac{d\textbf{r}_i}{dt},
\end{equation}
where $m$ is the mass of each particle and $\zeta$ represents the drag coefficient due to the surrounding solvent.

For the static structural analysis detailed in section \ref{topo_anal}, we study the effect of varying the initial quench rate between $10^{-2}$ and $10^{-6} \epsilon /k_B\tau_0$ at  constant volume fraction $\phi = 0.1$, and we study the effect of varying the volume fraction from $0.05$ to $0.15$ at a constant quench rate $\Gamma \approx 10^{-6} \epsilon /k_B\tau_0$. For the rheological measurements in section \ref{topo_rheo}, we exclusively study a system prepared with volume fraction $\phi = 0.1$ and quench rate $\Gamma \approx 10^{-5} \epsilon /k_B\tau_0$, which has been extensively studied in prior work \cite{colombo2014stress,donley2022time}.

\subsection{Rheological Simulations}

The rheological response is measured as in \cite{donley2022time} by imposing an oscillatory strain signal $\gamma(t)=\gamma_0 \sin \omega t$ in the $xy$-plane of the simulation box through
\begin{equation}\label{oscillatory}
 m\frac{d^2{\textbf{r}}_i}{dt^2}=-\nabla_{{\textbf{r}}_i}U - \zeta\left(\frac{d{\textbf{r}}_i}{dt}-\dot{\gamma}(t)y_i{\hat{\textbf{x}}}\right) 
\end{equation}
while updating the Lees-Edwards boundary conditions at every time step. Here $\gamma_0$ is the strain amplitude and $\hat{\textbf{x}}$ denotes the unit vector in the $x$-direction. In all the rheological measurements we used a Stokes-like drag with $m/\zeta=0.5\tau_0$, having verified that the results do not change qualitatively with further decreasing the $m/\zeta$ ratio.
 
To systematically study how the gels' nonlinear rheological responses depend on their microstructures, we perform large amplitude oscillatory shear (LAOS) using a rate-controlled deformation:
\begin{equation}\label{osc_def}
\dot{\gamma}(t)=\gamma_0\omega cos(\omega t).
\end{equation}
We specifically chose to perform an amplitude sweep by varying the amplitude over the range $0.001 < \gamma_0 < 10$ strain units and holding the frequency of oscillation at $\omega$ = 0.0025$\tau_0^{-1}$. This range of strain amplitudes spans from the linear viscoelastic regime through the point where deformations are large enough to yield the material. It is worth noting that, every amplitude in this amplitude sweep corresponds to a separate rheological test that starts from the same unperturbed gel configuration \cite{donley2022time}. This differs from most experimental amplitude sweep tests, where the different amplitudes are typically performed sequentially.

Under applied deformation, we compute the instantaneous shear stress $\sigma_{xy} (t)$. The stresses are computed from the interaction part of the global stress tensor using the standard virial equation \cite{thompson09} while neglecting other contributions (kinetic and viscous terms) as in previous studies \cite{Bouzid:2018BookChap,colombo2014stress,donley2022time}.

The complex viscoelastic modulus $G^*(\omega)$ is obtained from the Fourier transforms of the stress output ${\tilde{\sigma}}(\omega)$ and the strain input ${\tilde{\gamma}}(\omega)$ signals, as $G^*(\omega)={\tilde{\sigma}}(\omega)/{\tilde{\gamma}}(\omega)$, from which we compute the storage modulus $G^\prime(\omega)$ and the loss modulus $G^{\prime \prime}(\omega)$ defined respectively as the real and imaginary part of $G^*(\omega)$. The data used for these moduli is taken from cycles once the response has reached steady alternance \cite{donley2022time} (typically achieved after 4-6 cycles of deformation). Prior oscillatory shear and steady shear measurements \cite{bouzid2018network,colombo2014stress,donley2022time} indicate that the dynamic moduli are linear at small strain amplitudes ($\gamma_0$)  while the response becomes nonlinear at large $\gamma_0$ values.

For three specific stress amplitudes identified from the amplitude sweep ($0.1$, $0.35$, and $1.0$ strain units) additional oscillatory tests were run for ten cycles of deformation, with rheological data and structural configurations being extracted from the system 32 times per period for use in the topological analysis discussed in section \ref{topo_rheo}. 

\section*{Acknowledgements}

This work was supported by the U.S. Department of Energy, Office of Science, Advanced Scientific Computing Research, under contract number DE-AC02-06CH11357. Victor M. Zavala acknowledges partial support of the U.S. National Science Foundation through the University of Wisconsin Materials Research Science and Engineering Center (DMR-2309000). Gavin J. Donley and Emanuela Del Gado acknowledge support from the U. S. National Science Foundation (grant DMREF CBET—2118962).

\section*{Code and Data Availability}

All code for the topological data analysis is available via Zenodo at: \url{https://doi.org/10.5281/zenodo.10909693}.  Raw data for molecular simulations is available upon request. 

\bibliography{References}

\end{document}